\documentclass[10pt,prb,aps,twocolumn,superscriptaddress,floatfix,showpacs,nobalancelastpage,longbibliography,hyperref]{revtex4-1}

\usepackage{array,longtable}
\newcolumntype{L}{>{\tiny $}p{0.33\columnwidth}<{$}}
\newcolumntype{M}{>{\scriptsize $}p{0.33\columnwidth}<{$}}
\newcolumntype{N}{>{\scriptsize $}p{0.43\columnwidth}<{$}}
\usepackage{float}
\usepackage{graphicx}
\usepackage{graphics}
\usepackage{amsmath}
\usepackage{amssymb}
\usepackage{amsfonts}
\usepackage{dsfont}
\usepackage{braket}
\usepackage{color}
\usepackage{braket,slashed}
\usepackage[mathscr]{euscript}
\usepackage{hyperref}
\usepackage{epstopdf}
\usepackage{subfigure}
\usepackage{xfrac}
\usepackage{bm}
\usepackage{natbib}
\usepackage{kantlipsum}

\allowdisplaybreaks[1]

\newif\ifhyper
\hypertrue
\ifhyper
\hypersetup{
   citecolor = {green},
   colorlinks = {true}, % false
   urlcolor = {blue} % magenta
}
\fi

\newcommand{\beq}{\begin{equation}}
\newcommand{\eeq}{\end{equation}}
\newcommand{\beqa}{\begin{eqnarray}}
\newcommand{\eeqa}{\end{eqnarray}}

\begin{document}

\title{Systematic construction of spin liquids on the square lattice\\ from tensor networks with SU(2) symmetry}
\author{Matthieu Mambrini}\affiliation{Laboratoire de Physique Th\'eorique, C.N.R.S. and Universit\'e de Toulouse, 31062 Toulouse, France}
\author{Rom\'an Or\'us}
\affiliation{Institute of Physics, Johannes Gutenberg University, 55099 Mainz, Germany}
\author{Didier Poilblanc}
\affiliation{Laboratoire de Physique Th\'eorique, C.N.R.S. and Universit\'e de Toulouse, 31062 Toulouse, France}
\date{today}                                           % Activate to display a given date or no date

\begin{abstract}
We elaborate a simple classification scheme of all rank-5 SU(2)-spin rotational symmetric tensors according to 
i) the on-site physical spin-$S$, (ii) the local Hilbert space $V^{\otimes 4}$ of the four virtual (composite) spins attached to 
each site and (iii) the irreducible representations of the $C_{4v}$ point group of the square lattice. We apply our scheme to draw a complete list of all SU(2)-symmetric translationally and rotationally-invariant Projected Entangled Pair States (PEPS) with bond dimension $D\leqslant 6$. All known SU(2)-symmetric PEPS on the square lattice are recovered and simple generalizations are provided in some cases. More generally, to each of our symmetry class can be associated a $({\cal D}-1)$-dimensional 
manifold of spin liquids (potentially) preserving lattice symmetries and defined in terms of ${\cal D}$ independent tensors of a given bond dimension $D$. In addition, generic (low-dimensional) families of PEPS explicitly breaking either (i) particular point-group lattice symmetries (lattice nematics) or (ii) time reversal symmetry (chiral spin liquids) or (iii) SU(2)-spin rotation
symmetry down to $U(1)$ (spin nematics or N\'eel antiferromagnets) can also be constructed.
We apply this framework to search for new topological chiral spin liquids characterized by well-defined chiral edge modes, as revealed by their entanglement spectrum. In particular, we show how the symmetrization of a double-layer PEPS leads to a chiral topological state with a gapless edge described by a SU(2)$_2$ Wess-Zumino-Witten model. 

\end{abstract}

\maketitle

\section{Introduction}

The study of quantum many-body entanglement has provided many key insights into the structure of quantum states of matter. Low-energy states of quantum lattice systems obey typically the so-called ``area law" of the entanglement entropy \cite{Calabrese2009,Eisert2010,Cirac2012a}. As such, the area-law is a huge constraint on the classes of states that capture the relevant properties of matter at low energies. A more refined study has shown that, in fact, those states are captured by the so-called tensor network states, or simply ``tensor networks" \cite{Schuch2013b,Orus2013}. Such states obey naturally the area-law, and are at the basis of many theoretical and numerical developments in the study of quantum many-body systems and beyond \cite{Orus2014}. Examples of such states are, e.g., Matrix Product States (MPS)~\cite{Vidal2004a}, Projected Entangled Pair States (PEPS)~\cite{Verstraete2004b,Verstraete2006}, and the Multiscale Entanglement Renormalization Ansatz (MERA)~\cite{Vidal2007a}. These structures are, respectively, behind the so-called Density Matrix Renormalization Group algorithm (DMRG) for 1d systems~\cite{White1992}, PEPS-algorithms for 2d systems~\cite{Verstraete2008a}, and Entanglement Renormalization~\cite{Vidal2009a}. 

The description of quantum many-body states in terms of tensor networks has several advantages. Apart from naturally obeying the area-law (and therefore capturing the correct expected entanglement behavior), TN states can also be manipulated efficiently (either exactly or approximately). Another advantage is the fact that both lattice and internal symmetries can be naturally incorporated. For instance, a description in terms of symmetric tensors~\cite{Singh2010a,Schuch2010a} can lead to important computational 
advantages~\cite{Singh2010b,Singh2011,Singh2012,Weichselbaum2012,Singh2013}, 
and helps in the theoretical classification of phases of 
matter~\cite{Rispler2015}. Moreover, gauge symmetries can also be naturally incorporated~\cite{Haegeman2015}, hence offering a natural framework to describe lattice gauge theories~\cite{Zohar2016a,Zohar2016b}. 

In a seminal paper~\cite{Jiang2015}, S.~Jiang and Y.~Ran made the first attempt to organize PEPS into crude classes distinguished by short-range physics, related to the fractionalization of both on-site symmetries and space-group symmetries.  In their work, the authors introduced (quite generally) the notion of projective symmetry group (PSG) for PEPS, enabling
to deal {\it a priori} with gauge equivalence between tensors. Using lattice quantum numbers, the authors predicted a number of district classes for spin-$\frac{1}{2}$ spin liquids on the Kagome lattice. More recently, a similar framework was applied to classify (trivial) \hbox{spin-$1$} PEPS on the square lattice~\cite{Lee2016}. 

Our goal in this paper is to produce a simple classification scheme of all rank-5 SU(2)-spin rotational symmetric tensors. We characterize the tensors according to three criteria: i) the on-site physical spin-$S$, (ii) the local Hilbert space $V^{\otimes 4}$ of the four virtual (composite) spins attached to each site and (iii) the irreducible representations of the $C_{4v}$ point group of the square lattice. Using this scheme, we produce explicit expressions for all SU(2)-symmetric translationally and rotationally-invariant PEPS with bond dimension $D\leqslant 6$. As we shall see, one can recover all known SU(2)-symmetric PEPS on the square lattice as particular cases in our classification. Generically, to each of our symmetry class can be associated a $({\cal D}-1)$-dimensional manifold of spin liquids (potentially) preserving lattice symmetries and defined in terms of ${\cal D}$ independent tensors of a given bond dimension $D$. In addition, generic (low-dimensional) families of PEPS explicitly breaking particular point-group lattice symmetries (lattice nematics) and/or time reversal symmetry (chiral spin liquids~\cite{Kalmeyer1987,Schroeter2007}) can also be constructed. Finally, we apply this framework to search for new topological chiral spin liquids characterized by well-defined chiral edge modes, as revealed by their entanglement spectrum, and show how the symmetrization of a given double-layer PEPS leads to a chiral topological state with a gapless edge described by a SU(2)$_2$ Wess-Zumino-Witten (WZW) model~\cite{Witten1983}. 

The paper is organized as follows: in Sec.~\ref{sec2} we elaborate on the specifics of our 
classification~\footnote{The complete list of SU(2)-spin rotationally invariant PEPS with bond dimension $D\leqslant 6$ as well as the expressions of all the corresponding site tensors are provided in the Supplemental Material.}, show how many remarkable states of matter fit into it, explain the procedure to construct spin liquids, and point out the connection to previous work. In Sec.~\ref{sec3} we explain several attempts to obtain PEPS corresponding to higher-spin chiral topological quantum spin liquids. In particular, we show how a double-layer PEPS leads to a chiral topological state with a gapless edge described by a SU(2)$_2$ WZW model, which we characterize through its entanglement spectrum (ES). We also discuss how the PEPS tensor of such double-layer wavefunction can be expanded as a sum of ``fundamental" SU(2)-invariant PEPS tensors, which we also characterize. In Sec.~\ref{sec4} we wrap up our conclusions and outline several directions for future work. Finally, in Appendix~\ref{app1} we review the calculation of entanglement spectrum (ES) from 2d PEPS, and in Appendices~\ref{app2} and \ref{app3} we provide the coefficients for the PEPS tensors of several remarkable 
and simple states, and of all tensors entering the 
decomposition of the double-layer CSL, respectively. 

\section{Classification of SU(2)-symmetric PEPS on the square lattice}
\label{sec2}

\subsection{Construction}

\begin{figure}
\begin{center}
\includegraphics[width=8cm,angle=0]{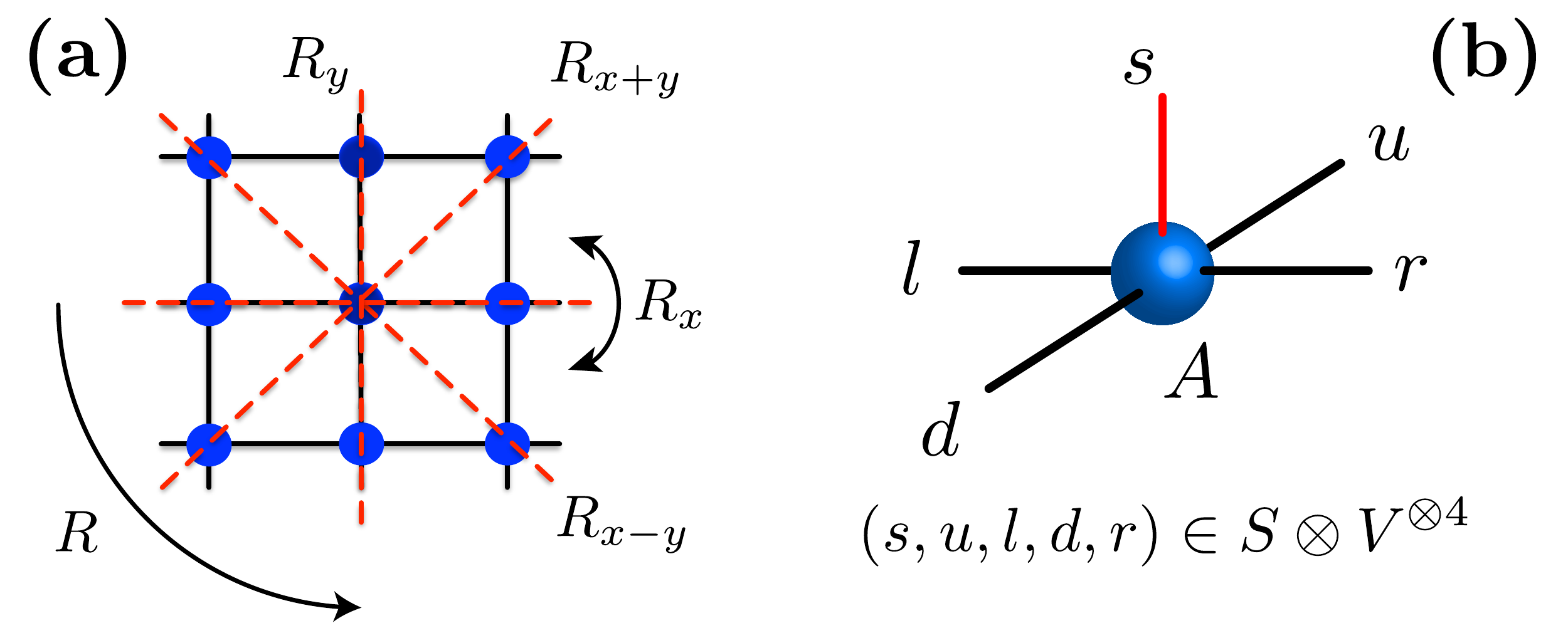}
\caption{[Color online] (a) Square lattice invariant under $C_{4v}$ point group symmetries (reflection axis are shown). 
The generators of the point group are, e.g., the 90-degree rotation $R$, the reflection $R_x$ and 
the inversion ${\cal I}=R_xR_y$. (b) A generic rank-5 PEPS tensor with one physical index $s$ and four virtual indices $u$, $l$, $d$ and $r$.}
\label{FIG:lattice_tensor}
\end{center}
\end{figure}

We consider here a square lattice (as shown in Fig.~\ref{FIG:lattice_tensor}(a)) with physical spin-$S$ site degrees of freedom. Hence, $d=2S+1$ basis states are assigned on each site. Here, we explicitly consider transitionally invariant states described by a PEPS built from a single tensor $A$, as the one shown in Fig.~\ref{FIG:lattice_tensor}(b). Each physical site has four virtual spins attached labelled on the figure by $u$, $l$, $d$ and $r$ along the up, left, down and right directions, respectively. 
The virtual states $|v_\alpha\big>$ belong to some representation $V$ of SU(2) of total dimension $D$ which, generically, 
is a direct sum $V=\oplus V_i$ of $\cal N$ irreducible representations (IRREP) $V_i$ of SU(2), each of partial dimension $2V_i+1$ 
and $D=\sum_{i=1}^{\cal N} (2V_i+1)$.

The corresponding translationally invariant PEPS~\cite{Cirac2012a} is obtained by assigning the same tensor $A$ on every physical site. Physically, the site tensor $A$ simply encodes a projector that maps the virtual space $V^{\otimes 4}$ onto all 
$2S+1$ components of the physical spin-$S$. From the bond point of view, every pair of the nearest-neighbor (NN) virtual spins is projected to a block diagonal {\it virtual spin singlet state}. By construction the obtained wave function is a global spin singlet, i.e. invariant under SU(2) rotations. For a bi-partite lattice as the square lattice one can perform a simple spin rotation (by $\pi$ around the $Y$-spin axis)
on all sites of a given sublattice that transforms the virtual bond singlets into diagonal maximally entangled NN pair states
$|{\cal S}\big> =\sum_{\alpha=1}^D |v_\alpha v_\alpha \big>$.  In this way the SU(2)-invariant PEPS becomes a simple 
contraction of the tensor network of the $A$'s. 

Our construction can also be easily generalized to states that are not SU(2)-singlets, i.e., have a total (average) spin component. For this, one could eventually build up tensors that transform under $S \neq 0$ IRREPS of SU(2) (sometimes dubbed "covariant" states). In this paper, however, we consider only "invariant" states under SU(2) -i.e. singlets-, which for simplicity we also call "symmetric".

The $2S+1$ components $A_s$ of a tensor $A$ encode the projectors $P_s : V^{\otimes 4}\rightarrow |s\big>$
onto the $|s\big>\equiv |S,S_z=s\big>$ physical state. Hence, the problem of enumerating all SU(2)-invariant PEPS 
reduces to the finding of all (orthogonal) projectors that map 
any virtual space $V^{\otimes 4}$ onto any spin-$S$ Hilbert space. Reversely, it amounts to enumerate
all $D^4$ orthogonal spin-$S$ wave functions that can be constructed out of the $(\oplus V_i)^{\otimes 4}$ basis
states.
In other words, we shall use the one-to-one correspondence between projectors and wave functions and
extract the tensor components from the wave functions.
One can write the physical state $|s\big>$ as
\begin{equation}
|s\big>=\sum_{\alpha_1,\alpha_2,\alpha_3,\alpha_4}
A_s(\alpha_1,\alpha_2,\alpha_3,\alpha_4) |\alpha_1,\alpha_2,\alpha_3,\alpha_4\big>, 
\end{equation}
 in terms of the $D^4$ virtual basis states
$|\alpha_1,\alpha_2,\alpha_3,\alpha_4\big>$ and of the tensor elements $A_s(\alpha_1,\alpha_2,\alpha_3,\alpha_4)$. 
Since we can always find a set of $D^4$ orthogonal wave functions, the corresponding basic tensors fulfill the ``orthonormalisation'' property
\begin{eqnarray}
%\sum_{\alpha_1}\sum_{\alpha_2}\sum_{\alpha_3}\sum_{\alpha_4} 
\sum_{\alpha_1,\alpha_2,\alpha_3,\alpha_4} [A_s(\alpha_1,\alpha_2,\alpha_3,\alpha_4)]^* B_{s'}(\alpha_1,\alpha_2,\alpha_3,\alpha_4) 
\nonumber \\  =\delta_{s s'} \delta_{AB}\, ,
\end{eqnarray}
where the left-hand side defines some tensor inner product $A_s\cdot B_{s'}$.
We have performed such a program {\it analytically} using Mathematica for all possible virtual spaces with
$D\leqslant 6$.  To reduce the cost of the computation, it is advantageous to use spin and (lattice) point group symmetries. First, it is convenient to decompose the virtual space $V^{\otimes 4}$ into all disconnected subspaces given 
by the occupations $n_{\rm occ.}=\{n_1,\cdots, n_{\cal N}\}$ of the $\cal N$ spins $V_i$ ($\sum_{i=1}^{\cal N} n_i=4$), each subspace providing a  different class of tensors. Secondly, we use the $S_z$ quantum number of the (physical) wave function. In fact, we start by computing all
$S_z=S$ wave functions (or equivalently all projectors onto the maximum $S_z=S$ subspace) and then apply the spin-lowering operator $S_-$ written in the virtual basis states. Simultaneously, we classify the various spin-$S$ wave functions according to
their point symmetry, i.e., according to the representations of the  $C_{4v}$ point group:  $A_1$ ($s$-wave), 
$B_1$ ($d_{x^2-y^2}$-wave), $E$ (doubly-degenerate $p$-wave), $A_2$ ($g$-wave) and $B_2$ ($d_{xy}$-wave).
To accomplish such a purpose, we simply need to diagonalize, in the space spanned by the
$(\oplus V_i)^{\otimes 4}$ virtual basis
attached to a given site, simultaneously i) the total spin operator, ii) the 90-degree rotation operator $R$, iii) the reflection symmetry
$R_x$ operator, and iv) the inversion ${\cal I}=R_xR_y$  operator (see Fig.~\ref{FIG:lattice_tensor}(a)). 
In practice, to perform this task efficiently, we have constructed in the $V^{\otimes 4}$ basis, the combined
(non-Hermitian) operator
\begin{equation}
{\cal O}_{\sigma,\sigma_z,\rho,\delta,\nu}=\sigma {\bf S}^2 + \sigma_z S_z + \rho R + \delta R_x + \nu D
\end{equation}
where the diagonal operator $D$ has specific diagonal elements characterizing each $n_{\rm occ.}$ sector.
The (real) coefficients $\sigma,\sigma_z,\rho,\delta,\nu$ are all chosen of very different magnitudes -- e.g. $10^{8}, 
10^{6}, 10^{4}, 10^{2}, 1$ -- in order to sort out the various eigenvalues (of order $1$). 
Note that tensors belonging
to the $A_1$, $B_1$, $A_2$ and $B_2$ symmetry classes are purely real while the 
$E$-symmetric tensors are intrinsically complex and come in complex conjugated pairs.

As a simple example, let us consider the case $V=\frac{1}{2}$ which contains 
$2^4=16$ basis states, all in the unique $n_{\rm occ.}=\{4\}$ sector ($V$ here is a simple IRREP).  
Diagonalizing ${\cal O}_{\sigma,\sigma_z,\rho,\delta,\nu}$ (omitting the $D$ part) gives 16 states (or tensor components) which can be grouped into
two singlets ($S=0$), three triplet ($S=1$) and one quintuplet ($S=2$), in agreement with the decomposition 
$(\frac{1}{2})^{\rm \otimes 4}=(0\oplus 1)^{\rm \otimes 2}=2(0)\oplus 3(1)\oplus (2)$. 
The $S=0$ outcomes correspond to ``classical'' TN, not considered afterwards. Inspection of the
eigenvalues associated to $R$ and $R_x$ reveals that one of the triplet states (or tensors) has $B_1$
symmetry, while the other two form a complex conjugate pair of  $E$ symmetry. On the other hand, 
the unique $S=2$ tensor is fully symmetric ($A_1$ IRREP) and corresponds to the spin-$2$ AKLT state
(see below).

\begin{table*} [htb]
\begin{center}
\resizebox{2.08\columnwidth}{!}{%
  \begin{tabular}{@{} ccccc@{}}
   \hline 
   \hline
    V $\backslash$ S & 1/2 & 1 & 3/2  & 2 \\ 
    \hline
    $\frac{1}{2}$&  & $\bf B_1$ $E$ & &  $\bf A_1$ \\
     \hline 
    $\frac{1}{2}\oplus 0$ & & & & \\
    $n_{\rm occ.}=\{1,3\}$ & $\bf A_1$[$\bf B_1$] $\bf E$ & & & \\
    $n_{\rm occ.}=\{2,2\}$ & & $\bf A_1^{(a)}$ $\bf A_1^{(b)}$[$\bf B_1$] $E$  $B_2$& & \\
    $n_{\rm occ.}=\{3,1\}$ & $\bf A_1$[$\bf B_1$] $E^{(a,b)}$ $\bf A_2$[$\bf B_2$]& &$A_1$[$B_1$] $E$ &\\
       \hline 
        $1$ &   & $B_1$ $E^{(a,b)}$ $A_2$& & $A_1^{(a)}$[$B_1$] $A_1^{(b)}$ $E$ $B_2$\\
       \hline 
        $\frac{1}{2}\oplus 0\oplus 0$ & & & &    \\
   $n_{\rm occ.}=\{1,1,2\}$ & $A_1^{(a,b)}$[$B_1^{(a,b)}$] $E^{(a-c)}$ $A_2$[$B_2$]& & & \\
   $n_{\rm occ.}=\{2,1,1\}$ & & $A_1^{(a,b)}$[$B_1^{(a,b)}$] $E^{(a-c)}$ $A_2$[$B_2$]& & \\
       \hline 
        $\frac{1}{2}\oplus \frac{1}{2}$ &  & & &   \\
    $n_{\rm occ.}=\{1,3\}$ & & $A_1^{(a,b)}$[$B_1^{(a,b)}$] $E^{(a-c)}$ $A_2$[$B_2$]& &$A_1$[$B_1$] $E$ \\
    $n_{\rm occ.}=\{2,2\}$ & & $A_1^{(a,b)}$[$B_1^{(a,b)}$] $B_1^{(c)}$ $E^{(a-e)}$ $A_2^{(a)}$ $A_2^{(b)}$[$B_2$] & 
    & $A_1^{(a)}$ $A_1^{(b)}$[$B_1$] $E$ $B_2$ \\
      \hline 
        $1\oplus 0$ &  & & & \\
    $n_{\rm occ.}=\{1,3\}$ & & $\bf A_1$[$\bf B_1$] $E$ & & \\
    $n_{\rm occ.}=\{2,2\}$ & &  $B_1$[$A_2$] $E^{(a,b)}$  & & $\bf A_1^{(a)}$[$\bf B_2$] $\bf A_1^{(b)}$[$\bf B_1$] $E$ \\
    $n_{\rm occ.}=\{3,1\}$ & & $A_1^{(a,b)}$[$B_1^{(a,b)}$] $E^{(a-c)}$ $A_2$[$B_2$] & &$A_1$[$B_1$] $E$ \\
       \hline 
   $\frac{3}{2}$&  & $B_1^{(a)}$ $B_1^{(b)}$ $E^{(a-c)}$ $A_2$ & &  $A_1^{(a-c)}$  $B_1$ $E^{(a)}$ $E^{(b)}$ $A_2$[$B_2^{(a)}$] $B_2^{(b)}$ \\
           \hline 
        $\frac{1}{2}\oplus 0\oplus 0\oplus 0$ &  & & &   \\
  $n_{\rm occ.}=\{1,1,1,1\}$ & $A_1^{(a-c)}$[$B_1^{(a-c)}$] $E^{(a-f)}$ $A_2^{(a,b)}$[$B_2^{(a,b)}$]& & & \\        
       \hline 
        $\frac{1}{2}\oplus \frac{1}{2}\oplus 0$ & & & &   \\
    $n_{\rm occ.}=\{1,2,1\}$  & $A_1^{(a-c)}$[$B_1^{(a-c)}$] $E^{(a-f)}$ $A_2^{(a-c)}$[$B_2^{(a-c)}$]& & $A_1^{(a,b)}$[$B_1^{(a,b)}$] $E^{(a-c)}$ 
    $A_2$[$B_2$] & \\        
    $n_{\rm occ.}=\{1,1,2\}$  & & $A_1^{(a,b)}$[$B_1^{(a,b)}$] $E^{(a-c)}$ $A_2$[$B_2$] & & \\        
      \hline 
        $1\oplus 0\oplus 0$ & & & &    \\
    $n_{\rm occ.}=\{1,1,2\}$ & & $A_1^{(a,b)}$[$B_1^{(a,b)}$] $E^{(a-c)}$ $A_2$[$B_2$] & & \\
    $n_{\rm occ.}=\{2,1,1\}$ & &  $A_1$[$B_1$] $E^{(a-c)}$  $A_2^{(a,b)}$[$B_2^{(a,b)}$]& & $A_1^{(a,b)}$[$B_1^{(a,b)}$]  $E^{(a-c)}$  $A_2$[$B_2$]\\
    \hline
       $1\oplus \frac{1}{2}$ &  & &  &  \\
 $n_{\rm occ.}=\{1,3\}$ & $A_1^{(a,b)}$[$B_1^{(a,b)}$] $E^{(a-c)}$ $A_2$[$B_2$] & 
 & $A_1^{(a,b)}$[$B_1^{(a,b)}$] $E^{(a-c)}$ $A_2$[$B_2$] & \\
    $n_{\rm occ.}=\{2,2\}$ & & $A_1^{(a-c)}$[$B_1^{(a-c)}$] $A_1^{(d,e)}$ $E^{(a-e)}$ $A_2^{(a,b)}$[$B_2^{(a,b)}$] $B_2^{(c,d)}$ &
    & $A_1^{(a,b)}$[$B_1^{(a,b)}$] $B_1^{(c)}$ $E^{(a-e)}$ $A_2$[$B_2$] $B_2$\\
    $n_{\rm occ.}=\{3,1\}$ & $A_1^{(a,b)}$[$B_1^{(a,b)}$] $E^{(a-d)}$ $A_2^{(a,b)}$[$B_2^{(a,b)}$] &  
    & $A_1^{(a-c)}$[$B_1^{(a-c)}$] $E^{(a-e)}$ $A_2^{(a,b)}$[$B_2^{(a,b)}$]  & \\
    \hline 
   $\frac{3}{2}\oplus 0$ & & & &\\
      $n_{\rm occ.}=\{1,3\}$ & & &$\bf A_1$[$\bf B_1$] $E$ & \\
    $n_{\rm occ.}=\{2,2\}$ & & $A_1^{(a)}$ $A_1^{(b)}$[$B_1$] $E$  $B_2$& & $B_1$[$A_2$] $E^{(a,b)}$ \\
        $n_{\rm occ.}=\{3,1\}$ & $A_1$[$B_1$] $E^{(a,b)}$ $A_2$[$B_2$]& & $A_1^{(a,b)}$[$B_1^{(a,b)}$] $E^{(a-d)}$ $A_2^{(a,b)}$[$B_2^{(a,b)}$]  & \\  
        \hline 
   $2$ &   & $B_1^{(a,b)}$[$A_2^{(a,b)}$] $E^{(a-d)}$ & & $A_1^{(a,b)}$[$B_1^{(a,b)}$] $A_1^{(c,d)}$ $E^{(a-c)}$ $A_2$[$B_2^{(a)}$] $B_2^{(b,c)}$\\
      \hline 
        \hline 
      \end{tabular}
     }
\caption{List of all SU(2)-symmetric basic rank-5 tensors for physical spin $\frac{1}{2}\leqslant S\leqslant 2$ and bond dimension $D\leqslant 5$.
The virtual (physical) spin degrees of freedom $V$ ($S$)  is displayed vertically (horizontally). 
Whenever the virtual spin is a direct sum $V=\oplus V_i$, we decompose the virtual space $V^{\otimes 4}$ into 
all subspaces given by the occupations $n_{\rm occ.}$ of the spins $V_i$.  Note that, when two tensor classes are related by
``color exchange'', we keep only one of them in the list (e.g. for $V=\frac{1}{2}\oplus 0\oplus 0$ 
$n_{\rm occ.}=\{1,2,1\}$ is omitted since it is the ``color conjugate" of $n_{\rm occ.}=\{1,1,2\}$). Tensors are labelled according to their $C_{4v}$ point symmetry,  $A_1$ ($s$-wave), $B_1$ ($d_{x^2-y^2}$-wave), 
$E$ (doubly-degenerate $p$-wave),
$A_2$ ($g$-wave), $B_2$ ($d_{xy}$-wave). Subscripts $(a)$, $(b)$, etc... are used to differentiate non-equivalent tensors of the same class. Gauge-equivalent tensors (i.e. giving rise to the same PEPS) are listed between brackets, next to their gauge-related partners. Tensors giving rise to simple known wave functions
are highlighted in boldface (see text for a description and Supplemental Material for expressions).
}
\label{Table:tensor_list}
\end{center}
\end{table*}

Using the method described above, we have generated all SU(2)-invariant tensors up to $D=6$.
A subset of the list of tensors with $D\leqslant 5$ and $\frac{1}{2}\leqslant S\leqslant 2$ is displayed in Table~\ref{Table:tensor_list}, 
while the complete list for $D\leqslant 6$ is given as Supplemental Material. The columns correspond to different physical spins (limited to $S\leqslant 2$). 
Vertically, the tensors are classified according to the representation $V=\oplus V_i$ of the virtual spins.
Each line corresponds to a sub-class given by a disconnected set of (virtual) basis states
characterized by the occupation numbers $n_{\rm occ.}$ of the $\cal N$ spins $V_i$.

It should be mentioned that all the tensors produced in our systematic procedure do not lead 
necessarily to different PEPS due to remaining gauge degrees of freedom.
Indeed, imposing SU(2) and point group symmetries does not completely fix the gauge and some
 freedom remains. Typically, tensors which have identical non-zero tensor elements up to a sign
 are gauge equivalent as can be checked case by case. Table~\ref{Table:tensor_list} shows explicitly 
 all gauge-equivalent tensors. 
\begin{table}[htb]
\begin{center}
\resizebox{0.98\columnwidth}{!}{%
  \begin{tabular}{@{} ccccc @{}}
\hline
    \hline 
    V $\backslash$ S & 1/2 & 1 & 3/2  & 2 \\ 
  \hline 
   & & & &\\
    $\frac{1}{2}$& 0/0/0/0/0  & 0/1/0/0/1 & 0/0/0/0/0 &  1/0/0/0/0 \\
    & & & & \\
     \hline 
     & & & &\\
    $\frac{1}{2}\oplus 0$ & 2/2/1/1/3  & 2/2/0/1/2 & 1/1/0/0/1 & 1/0/0/0/0  \\
   & & & &\\
        $1$ & 0/0/0/0/0  & 0/1/1/0/2 & 0/0/0/0/0 & 2/1/0/1/1 \\
  & & & & \\
     \hline 
     & & & &\\
        $\frac{1}{2}\oplus 0\oplus 0$ &  8/8/4/4/12 & 6/5/1/3/6 & 2/2/0/0/2 & 1/0/0/0/0  \\
& & & &\\
        $\frac{1}{2}\oplus \frac{1}{2}$ &  0/0/0/0/0 & 6/9/4/3/13 & 0/0/0/0/0 & 6/3/0/1/3   \\
& & & &\\
        $1\oplus 0$ &    0/0/0/0/0 & 3/5/3/1/8 & 0/0/0/0/0 & 5/3/1/3/4   \\
  & & & &\\
     $\frac{3}{2}$& 0/0/0/0/0 & 0/2/1/0/3 & 0/0/0/0/0 & 3/1/1/2/2  \\
     & & & &\\
     \hline 
     & & & &\\
        $\frac{1}{2}\oplus 0\oplus 0\oplus 0$ &   21/21/12/12/33 & 12/10/3/6/13 & 3/3/0/0/3 & 1/0/0/0/0    \\    
 & & & &\\
        $\frac{1}{2}\oplus \frac{1}{2}\oplus 0$ &   10/10/8/8/18 & 12/13/5/6/18 & 6/6/2/2/8 & 6/3/0/1/3   \\
   & & & &\\
        $1\oplus 0\oplus 0$ & 0/0/0/0/0 & 11/14/9/6/23 & 0/0/0/0/0 & 10/7/3/6/10       \\
   & & & &\\
        $1\oplus \frac{1}{2}$ &    4/4/3/3/7 & 5/5/3/4/8 & 5/5/3/3/8 & 5/4/2/2/6    \\
   & & & &\\
   $\frac{3}{2}\oplus 0$ & 1/1/1/1/2 & 2/3/1/1/4 & 3/3/2/2/5 & 3/2/2/2/4 \\
   & & & &\\
   $2$ &  0/0/0/0/0 & 0/2/2/0/4 & 0/0/0/0/0 & 4/2/1/3/3 \\
  & & & &\\
     \hline 
    & & & &\\
         $\frac{1}{2}\oplus 0\oplus 0\oplus 0\oplus 0$ &   44/44/28/28/72 & 20/17/6/10/23 & 4/4/0/0/4 & 1/0/0/0/0    \\    
 & & & &\\
        $\frac{1}{2}\oplus \frac{1}{2}\oplus 0\oplus 0$ &   28/28/20/20/48 & 25/24/11/14/35 & 12/12/4/4/16 & 6/3/0/1/3  \\
  & & & &\\
        $\frac{1}{2}\oplus \frac{1}{2}\oplus \frac{1}{2}$ &  0/0/0/0/0 & 33/39/24/21/63 & 0/0/0/0/0 & 21/15/3/6/18   \\
    & & & &\\
        $1\oplus 0\oplus 0\oplus 0$ &    0/0/0/0/0 & 27/31/22/18/53 & 0/0/0/0/0 & 17/13/6/10/19    \\
    & & & &\\
        $1\oplus \frac{1}{2} \oplus 0$ &   11/11/8/8/19 & 13/13/8/9/21 & 11/11/7/7/18 & 10/8/4/5/12     \\
   & & & &\\
        $1\oplus 1$ &   0/0/0/0/0 & 9/13/13/9/26 & 0/0/0/0/0 & 19/15/7/11/22     \\
& & & &\\
   $\frac{3}{2}\oplus 0\oplus 0$ &  2/2/2/2/4 & 6/6/2/3/8 & 10/10/6/6/16 & 4/4/5/4/9 \\
     & & & &\\
   $\frac{3}{2}\oplus \frac{1}{2}$ & 0/0/0/0/0 & 7/12/8/5/20 & 0/0/0/0/0 & 15/10/7/10/17 \\
    & & & &\\
   $2\oplus 0$ & 0/0/0/0/0 & 1/4/5/2/9 & 0/0/0/0/0 & 10/7/3/6/10 \\
& & & &\\
  $\frac{5}{2}$ &  0/0/0/0/0 & 0/3/2/0/5 & 0/0/0/0/0 & 5/2/2/4/4 \\
& & & &\\
    \hline 
      \end{tabular}
     }
\caption{Sets of the numbers ${\cal D}_1/{\cal D}_2 / {\cal D}_3 / {\cal D}_4 / {\cal D}_5$ of basic tensors belonging to the five $A_1$, $B_1$, $A_2$, $B_2$ and (doubly degenerate) $E$ IRREP of $C_{4v}$, respectively, which can be combined
in each $SU(2)$ $(V,S)$ symmetry class to give rise to fully-symmetric spin-$S$ spin liquids, for $D\leqslant 6$ and $\frac{1}{2}\leqslant S\leqslant 2$.  Note each class defined by a direct sum $V=\oplus V_i$ of $\cal N$  SU(2) IRREP 
includes all basic tensors of all sub-classes defined by $n_{\rm occ.}=(n_1,n_2,\cdots n_{\cal N})$ with $0\leqslant n_i\leqslant 4$ and $\sum_i n_i=4$. For instance, the $V=\frac{1}{2}\oplus \frac{1}{2}\oplus 0$ ($D=5$) class includes the $V=\frac{1}{2}$ ($D=2$), $V=\frac{1}{2}\oplus 0$ ($D=3$) and 
$V=\frac{1}{2}\oplus \frac{1}{2}$ ($D=4$) classes. 
}
\label{Table:numbers}
\end{center}
\end{table}

\subsection{Remarkable PEPS in this scheme}

Some of the low-$D$ PEPS obtained using our systematic construction and 
listed in Table~\ref{Table:tensor_list} have already been introduced 
in recent literature and/or correspond to well-known states of matter. Let us briefly list a few of them below, together 
with some simple generalizations (see tensor expressions in Appendix~\ref{app2}). 

\subsubsection{The spin-$S$, $S$ even integer, AKLT states} 

The simplest 2d Affleck-Kennedy-Lieb-Tasaki (AKLT)~\cite{Affleck1987} state is obtained by decomposing a physical spin-2 on each site into four virtual 
spin-$\frac{1}{2}$ spins, pairing every NN virtual spins into singlet and, finally, applying the projector $P: \frac{1}{2}^{\otimes 4}\rightarrow 2$
onto the fully symmetric $S=2$ subspace on every site. This state is known to have short-range correlations, and is given by the $A_1$ tensor of the $V=\frac{1}{2}$ line / $S=2$ column of Table~\ref{Table:tensor_list}. This construction can be straightforwardly
extended to higher physical spins by projecting four virtual spin-$\frac{k}{2}$, $k\in \mathbb{N}^*$, onto the fully symmetric spin-$S$, $S=2k$, physical subspace. This is realized by the unique $V=\frac{k}{2}$, $S=2k$, $A_1$ tensor 
of dimension $D=k+1$ (see Appendix~\ref{app2} for $k=1$ and Supplemental Material for $k=1,\dots,5$). The 2d spin-$S$ AKLT states can serve as useful examples of trivial (featureless) states or 
states with symmetry protected topological order~\cite{Chen2011} for $k$ even integer ($S=4p$) or $k$ odd integer
($S=4p+2$), respectively.

\subsubsection{The spin-$S$, $S$ odd integer, featureless paramagnets}

Starting with four $V=\frac{1}{2}$ virtual spins attached to each site and paired up into NN singlets, as in the AKLT construction, but projecting them onto $S=1$ on-site physical spins gives rise to a spin-$1$ featureless paramagnet~\cite{Jian2016}, also with short-range correlations. This is given by the $V=\frac{1}{2}$, $S=1$, $B_1$ tensor of Table~\ref{Table:tensor_list}. This construction can be 
straightforwardly generalized to higher physical spins by simply
projecting four virtual spin-$\frac{k}{2}$, $k\in{\mathbb N}^*$, onto spin-$S$, $S=2k-1$, physical subspace. 
This is always realized by the unique $V=\frac{k}{2}$, $S=2k-1$, $B_1$ tensor
of dimension $D=k+1$ (see Appendix~\ref{app2} for $k=1$ and Supplemental Material for $k=1,\dots,5$). We believe such states with $S>1$ are also featureless paramagnets.
Note that, for all $k$, the $B_1$ tensor comes always in pair with a (complex) $E$ tensor which leads to a {\it real} wave function and 
might also have interesting properties. 

\subsubsection{The spin-$\frac{1}{2}$ Resonating Valence Bond (RVB) state} 

The RVB state is a  spin-1/2 spin liquid 
defined by an equal-weight superposition of all nearest-neighbor (NN) singlet configurations~\cite{Anderson1973}.  
It is exactly given by the $V=\frac{1}{2}\oplus 0$, $n_{\rm occ.}=\{1,3\}$, $A_1$ (named ${\cal A}_1^{(1)}$ from now on)
or $B_1$ tensors (named ${\cal B}_1^{(1)}$ from now on) of the $S=\frac{1}{2}$ column of
Table~\ref{Table:tensor_list}. The NN RVB state was shown to be a ${\mathbb Z}_2$ topological spin liquid on the Kagome lattice~\cite{Schuch2012,Poilblanc2012,Poilblanc2013a}. On the square lattice it exhibits an extended $U(1)$ gauge symmetry and is critical~\cite{Albuquerque2010,Poilblanc2012}
(see below for the discussion of gauge symmetry). 

\subsubsection{The long-range spin-$\frac{1}{2}$ RVB state (LR RVB)} 

The LR RVB state is obtained by assuming a distribution of longer-range singlet bonds beyond NN
(yet still connecting two different sublattices). It is obtained within the PEPS formalism 
by linearly combining the $V=\frac{1}{2}\oplus 0$, $n_{\rm occ.}=\{3,1\}$, $A_1$ (named ${\cal A}_1^{(2)}$ from now on) 
tensor with the previous ${\cal A}_1^{(1)}$ NN RVB tensor. 
\begin{equation}
{\cal A}_{\rm LRRVB}=\lambda_1 {\cal A}_1^{(1)} + \lambda_2 {\cal A}_1^{(2)},
\end{equation}
with $\lambda_1,\lambda_2\in \mathbb{R}$.
Alternatively, one can use the gauge-equivalent $B_1$ tensors, named ${\cal B}_1^{(1)}$ and ${\cal B}_1^{(2)}$.
The singlet bond distribution is controlled by the relative (real) 
weight between the two tensors. Such a spin liquid ansatz turned out to be an excellent variational state for the frustrated 
$J_1$--$J_2$ antiferromagnetic Heisenberg model on the square lattice~\cite{Wang2013}. It is interesting to notice that 
this ansatz is in fact the most general $D=3$ SU(2)-invariant PEPS.

\subsubsection{The NN fermionic spin-$\frac{1}{2}$ RVB (NN fRVB)}
 
The NN fermionic-RVB (fRVB) is defined as an equal-weight superposition of dimer coverings where each (centro-symmetric) dimer is 
written in the fermionic representation.  It can be re-written as a spin-$\frac{1}{2}$ NN RVB state where, e.g., vertical dimers are assigned a 
complex factor $i$ providing a completely different sign structure~\cite{Kotliar1988} than the above NN RVB state.  This (real) wave function is given by the unique
$S=\frac{1}{2}$, $V=\frac{1}{2}\oplus 0$, $n_{\rm occ.}=\{1,3\}$, complex E tensor~\cite{Poilblanc2014}.

\subsubsection{The generalized spin-$S$ NN RVB} 

The spin-$1$ RVB state can be obtained by  attaching a single $S=1$ 
virtual spin on every site (accompanied by 3 spin-$0$). All NN virtual spins $1/2$ are again paired up into singlets which resonate. 
The $S=1$, $V=1\oplus 0$, $n_{\rm occ.}=\{1,3\}$ $A_1$ (or $B_1$) tensor corresponds exactly to such a spin liquid. 
This scheme can be generalized to any \hbox{spin-$S$} NN RVB and is always described by a single $V=S\oplus 0$, $n_{\rm occ.}=\{1,3\}$ $A_1$ (or $B_1$) tensor. The cases corresponding to $S=1$ and $S=\frac{3}{2}$ are highlighted in Table~\ref{Table:tensor_list} and the corresponding tensors are given in Appendix~\ref{app2}
(see Supplemental Material for all physical spin $S$ up to $S=2$).

\subsubsection{The spin-$S$ Resonating AKLT Loop (RAL) state} 

As shown by Li et al.~\cite{Li2014}, the spin-$1$ RAL state involves two virtual spin-$\frac{1}{2}$ and two virtual spin-$0$ attached to every site, i.e., 
the virtual subspace is $V=\frac{1}{2}\oplus 0$, $n_{\rm occ.}=\{2,2\}$. Physically, NN virtual spin-$1/2$ are paired up into singlets 
(as in the RVB state) and all virtual spins are then projected locally onto physical spins $1$ to produce AKLT loops. The 
two $S=1$, $V=\frac{1}{2}\oplus 0$, 
$n_{\rm occ.}=\{2,2\}$, $A_1^{(a)}$ and $A_1^{(b)}$ tensors encode the two possible site configurations of the loops with
180-degree or 90-degree angles. The RAL state on the square lattice is critical since the dimer-dimer correlations decay as a power law~\cite{Li2014}. Since the 1d AKLT chain can be extended to higher physical spin (see above for the 2d case), it is easy to generalize the RAL to a gas of resonating spin-$S$ AKLT chains, for all $S$ integer. It is given by the only two $V=\frac{S}{2}\oplus 0$ ($D=S+2$), $n_{\rm occ.}=\{2,2\}$, $A_1^{(a)}$ and $A_1^{(b)}$ tensors.  The cases corresponding to $S=1$ and $S=2$ are highlighted in Table~\ref{Table:tensor_list} and the corresponding $S=1$ tensors are given in Appendix~\ref{app2}
(see Supplemental Material for all integer physical spin $S$ up to $S=4$). 

\subsubsection{The spin-$1/2$ chiral spin liquid (CSL)} 

\begin{figure}[htbp]
\begin{center}
\includegraphics[width=0.95\columnwidth,angle=0]{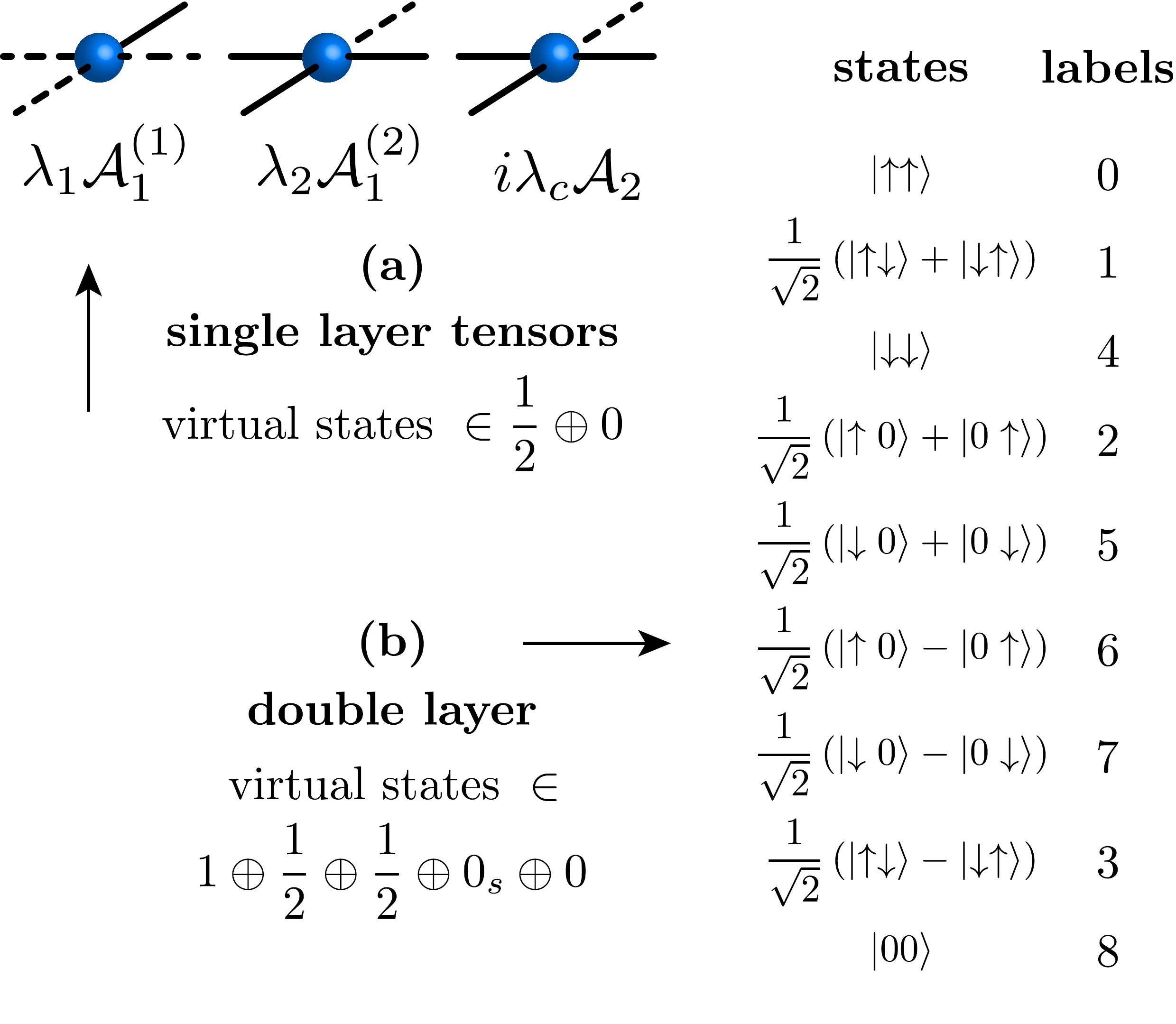}
\caption{(a) The three tensors of the spin-1/2 CSL involving the three virtual states of the $\frac{1}{2}\oplus 0$ (spin) representation, 
$|\uparrow\big>$ and $|\downarrow\big>$ on the full lines, and $|0\big>$ on the dotted lines. (b) The natural basis of virtual states of the double layer tensor $|\uparrow\uparrow\big>$, $|\uparrow\downarrow\big>$,  $|\downarrow\uparrow\big>$ and $|\downarrow\downarrow\big>$ 
($\in \frac{1}{2}\otimes\frac{1}{2}$, spin-$\frac{1}{2}$ on top {\it and} bottom layers), 
$|\uparrow 0\big>$, $|\downarrow 0\big>$, $|0\uparrow\big>$ and $|0\downarrow\big>$ 
($\in \frac{1}{2}\oplus\frac{1}{2}$, spin-$\frac{1}{2}$ on top {\it or} bottom layer), 
and $|00\big>$ ($\in 0$) is transformed (by a simple unitary transformation) into symmetric/antisymmetric states w.r.t to the exchange of layers. }
\label{FIG:single_double}
\end{center}
\end{figure}

The CSL is obtained by linearly combining the two previous ${\cal A}_1^{(1)}$ and ${\cal A}_1^{(2)}$
(with real coefficients) and the $V=\frac{1}{2}\oplus 0$, $n_{\rm occ.}=\{3,1\}$, $A_2$ tensor (named  ${\cal A}_2$ from now on)
with a pure-imaginary coefficient~\cite{Poilblanc2015,Poilblanc2016}. The PEPS obtained from the resulting tensor (see Fig.~\ref{FIG:single_double}(a)),
\begin{equation}
{\cal A}_{\rm chiral}=\lambda_1 {\cal A}_1^{(1)} + \lambda_2 {\cal A}_1^{(2)} + i \lambda_c {\cal A}_2,
\label{Eq:chiral_sl}
\end{equation}
with $\lambda_1,\lambda_2, \lambda_c\in \mathbb{R}$, breaks time-reversal symmetry (provided 
$\lambda_2 \lambda_c\ne 0$) and transforms into its complex conjugate state under
any of the reflection symmetries of Fig.~\ref{FIG:lattice_tensor}(a). It exhibits clear SU(2)$_1$ edge modes although there are some evidence for critical (singlet) bulk correlations. 

\subsection{Constructing spin liquids and beyond} 

\subsubsection{Generic spin liquids}

From a few of the previous examples, we see that tensors can be linearly combined to give new interesting states. In fact,
adding $\cal D$ (real) tensors $T^{(a)}$ as $\sum_{a=1}^{\cal D} \lambda_a T^{(a)}$ (involving $\cal D$ real coefficients $\lambda_a$)
belonging to the same ``class", i.e., characterized by the
same physical ($S$) and virtual ($V$) degrees of freedom and by the same IRREP of the point group $C_{4v}$, will lead to a $({\cal D}-1)$-dimensional
family of completely symmetric spin liquids which (potentially) do not break any symmetry, neither SU(2) nor point group symmetries. 
The numbers $\cal D$ of tensors which can be combined
in each SU(2) $(V,S)$ symmetry class to give rise to fully-symmetric spin-$S$ spin liquids are given in Table~\ref{Table:numbers} for $D\leqslant 6$ and $S\leqslant 2$. Note that the counting of tensors of a given bond dimension includes all those of smaller bond dimensions which can be combined in each class. 
Note also that, for a given linear combination of tensors, there is {\it a priori} no guarantee that all correlations, remain short range in such a symmetric state and the absence of spontaneous
symmetry breaking in the thermodynamic limit should, in principle, always be verified. We observe that the typical dimensions of the 
PEPS families do not grow too fast, from ${\cal D}\sim 3-10$ for $D=4$ up to ${\cal D}\sim 10-40$ for $D=6$.
Also, it is interesting to notice that a subset of the symmetry classes does not provide a variational representation of half-integer spin-$S$.

\subsubsection{Lattice nematics}

If the $\cal D$ $T^{(a)}$ tensors belong to (at least) two different IRREP of the point group (while still involving the same virtual
and physical degrees of freedom), the resulting PEPS will explicitly
break the point group symmetry. For example, combining $A_1$ and $B_1$ tensors, or  $A_2$ and $B_2$ tensors, will
produce a nematic state where vertical and horizontal directions will become non-equivalent (e.g. observables will acquire different 
mean values). 
As a concrete example, let us consider the $V=1$, $S=2$, gauge-equivalent $A_1^{(a)}$ and $B_1$ tensors of Table~\ref{Table:tensor_list}.
It is likely that these tensors produce a paramagnet similar to the $S=2$ AKLT state, although with gapped edge states. 
The linear combination $A_1^{(a)}+B_1$ ($A_1^{(a)}-B_1$) of the two tensors gives a product of decoupled vertical (horizontal) spin-$2$ AKLT {\it chains} times a collection of independent horizontal (vertical) NN dimers (constructed from pairs of all the remaining virtual spin-$1$ not involved in the chains). Any partial superposition
like $\cos{\theta}\, A_1^{(a)}+\sin{\theta}\, B_1$, $\theta\in ]0,\pi/2[$, will give a lattice nematic state interpolating between the isotropic paramagnet and the array of AKLT chains.

\subsubsection{Breaking SU(2)-symmetry down to U(1): spin nematics and N\'eel antiferromagnets}

By breaking the global SU(2)-spin rotation invariance down to U(1) 
one can construct, within our framework, two enlarged families of anisotropic quantum magnets. If TR-symmetry and space group symmetry are independently conserved, one obtains  
anisotropic spin nematics~\cite{Andreev1984} for which 
the spin Z-axis becomes non-equivalent from the two equivalent X and Y spin axis. If the combination of TR 
with a unit translation is conserved, one gets N\'eel-like quantum magnets with a finite staggered magnetization.
In order to achieve this goal, one should remember that our  
PEPS are defined in a physical basis where all spins on the B-sublattice have been rotated by $\pi$. 
Each $S_z$-component (over the $d=2S+1$ components) of a given tensor hence contributes to a finite amplitude of the (physical) staggered magnetization ${\tilde S}_z^{\rm stag}=S_z$ in the original un-rotated basis.  

The procedure to construct anisotropic magnets is therefore simple~; (i) One sorts out all tensors according to the virtual space $V$ (as before) and to 
some maximum value $S_{\rm max}$ 
of $|S_z|$, defining the physical Hilbert space $S_z\in [-S_{\rm max},S_{\rm max}]$, by merging classes of different SU(2)-spin $S$
with the same orbital symmetry (i.e. IRREP of $C_{4v}$); (ii) One groups all tensors into $S_z=\pm |S_z|$ pairs of tensors 
$T_{\pm}^{(a)}$ (where $a$ labels the pairs)
related by the $S_z\leftrightarrow -S_z$ symmetry (for $S_z\ne 0$);  
(iii) On then constructs the linear superposition of the (normalized) on-site tensors for each local $S_z=s|S_z|$ physical index as
$\sum_{a,s} \lambda_{a,s} T_{s}^{(a)}$.
Generically, such a family of PEPS exhibits a finite staggered magnetization (in the original un-rotated basis)
as in a N\'eel state unless we further impose $\lambda_{a,s}=\lambda_{a,-s}$ 
to construct nematic states. 
Note that Long Range Order (LRO) in the XY (spin) plane may still spontaneously appear in some domains 
of the parameter space, despite 
the local $U(1)$ symmetry. 

\subsubsection{Complex $E$ tensors, TR-symmetry breaking and chiral spin liquids}

It is important to notice that, if the $\cal D$ tensors belong to the $A_1$, $B_1$, $A_2$ or $B_2$ IRREP,  the resulting tensor is real
and, hence, invariant under time-reversal (TR) symmetry.  For complex coefficients, or when $E$ tensors are combined, the resulting state generically
breaks TR, except at fine-tuned parameter subsets. Although chiral spin liquids with protected chiral edge modes have to be searched in 
these classes of PEPS, we believe they probably span a tiny fraction of the TR-symmetry breaking PEPS manifold. Note also that not all $E$ tensors
give complex PEPS. For example, combining the three $S=1/2$, $V=\frac{1}{2}\oplus 0$, complex $E$ tensors (one tensor with $n_{\rm occ.}=\{1,3\}$ and two tensors with $n_{\rm occ.}=\{3,1\}$) surprisingly gives a real (SU(2)-symmetric) wave function. We believe this PEPS family can be viewed physically as an extension of the NN fRVB (see above)
in terms of fRVB states with longer-range (fermionic) dimers.  
Similarly, the two PEPS given by the unique $S=1$, $V=\frac{1}{2}\oplus 0$, $E$ tensor and by the unique $S=3/2$, $V=\frac{1}{2}\oplus 0$, $E$ tensor are also real. Hence, we believe that the spin-$\frac{1}{2}$, spin-$1$ and spin-$\frac{3}{2}$ PEPS originated
from the $V=\frac{1}{2}\oplus 0$, $E$ tensors can probably all be mapped to real fermionic PEPS (fPEPS). 

\subsubsection{Gauge symmetry and topological order}

Whether or not a PEPS of a given family exhibits topological order is a rather subtle issue. The existence of a gauge symmetry, i.e.,   an Invariant Gauge Group (IGG),  plays a crucial role and is often a necessary condition. The AKLT states and the featureless paramagnets above have simple ${\cal N}=1$ virtual spaces with a single spin species in the four directions, $n_{\rm occ.}=4$. This is connected to ${\rm IGG}=\mathbb{I}$, characteristic of topologically {\it trivial} states.
 
Gauge symmetry (like ${\mathbb Z}_2$ or U(1)) can also be present, depending of the different $n_{\rm occ.}$ sectors involved in the construction 
of the variational manifold. For instance, the NN RVB state is defined by a unit tensor with $V=\frac{1}{2}\oplus 0$ and $n_{\rm occ.}=\{1,3\}$ which, 
in practice, implies that one and only one singlet dimer is attached to every site~\cite{Schuch2012}. This local constraint implies that the {\it number} of dimers cut by a line winding around an infinite cylinder is conserved, hence providing an infinite number of topological sectors associated to 
a U(1) gauge symmetry~\cite{Poilblanc2012}. On the other hand, the LR RVB state mixes $n_{\rm occ.}=\{1,3\}$ 
and $n_{\rm occ.}=\{3,1\}$ tensors so that, then, only the {\it parity} of the number of dimers cut by a circumference is conserved,
hence reducing the number of sectors to two and the gauge symmetry to $\mathbb{Z}_2$. 
More generally, the gauge symmetry can usually be inferred from the set of numbers of occupation 
$[n_{\rm occ.}^1,n_{\rm occ.}^2,\cdots, n_{\rm occ.}^{\cal D}]$
of the $\cal D$ superposed tensors. In general, one can always use a {\it minimal} global virtual basis
$V=\oplus V_i$, direct sum of $\cal N$ IRREP $V_i$ of SU(2), for which all the occupation
numbers $n_{\rm occ.}^j$ are given by sets of $\cal N$ numbers, i.e. 
$n_{\rm occ.}^j=\{n_1^j,\cdots, n_{\cal N}^j\}$ and $\sum_i n_i^j=4, \forall j$.  Subsequently, topological order 
can be characterized from the symmetry~\cite{Schuch2010a}.
However, (i) an extended gauge symmetry can emerge in the thermodynamic limit, as e.g. the U(1) symmetry in the case of the 
$\mathbb{Z}_2$ LR RVB
state;  (ii) Reversely, a mechanism of ``confinement" can suppress the topological 
order associated to the underlying gauge symmetry~\cite{Schuch2013a}. In any case, topological order can always be inferred from a thorough
investigation of the transfer operator~\cite{Schuch2013a}, and not only from the local symmetry of the tensor. 

\subsection{Connection with previous work}

Based on the original framework introduced by S.~Jiang and Y.~Ran~\cite{Jiang2015} and the notion
of projective symmetry group, Lee and Han provided a classification of (trivial) spin-$1$ PEPS on the square 
lattice~\cite{Lee2016}, in terms of lattice quantum numbers. 
In our classification scheme, we have recovered their results which correspond to a subset of our $S=1$
tensors~: the $V=\frac{1}{2}$ ($D=2$), $B_1$ tensor, the $V=1$ ($D=3$), $B_1$ and $A_2$ tensors, the 
$V=1\oplus 0$ ($D=4$), $B_1$, $n_{\rm occ.}=\{1,3\}$ (RVB state) and $n_{\rm occ.}=\{2,2\}$ (RAL state) tensors.  
While each tensor has (emergent) $U(1)$ IGG symmetry, a linear combination of them eliminates this gauge symmetry and produces a trivial state. 
Note that these authors did not report about neither the $S=1$, $E$ tensors nor the $S=1$, $V=1\oplus 0$ ($D=4$), $n_{\rm occ.}=\{3,1\}$ tensors enumerated in Table~\ref{Table:tensor_list}.
Combining, e.g., the  $S=1$, $V=1\oplus 0$ ($D=4$), $n_{\rm occ.}=\{1,3\}$ and $n_{\rm occ.}=\{3,1\}$  tensors
preserves a $\mathbb{Z}_2$ IGG and may lead to a topological spin liquid.

\section{Application: search for higher-spin ($S>\frac{1}{2}$) CSL}
\label{sec3}

One of the applications of our classification is the systematic construction of CSL
beyond the physical $S=\frac{1}{2}$ and virtual $V=\frac{1}{2}\oplus 0$ IRREPs, already realized.  Our goal is to construct a family of TR symmetry-breaking PEPS with linear dispersing chiral edge modes described by a CFT beyond SU(2)$_1$. To characterize the edge modes, we analyze the entanglement spectrum (ES) of the corresponding PEPS when wrapped around an infinite cylinder and splitted in two parts (left and right), as explained in Appendix \ref{app1}. We describe below several natural routes to find such CSL. 

\subsection{The complex $E$ tensors }

One of the first natural candidates would be the above $E$ tensors which are intrinsically complex. 
Therefore, their corresponding PEPS can potentially break TR symmetry and, hence, be relevant 
topological CSL. 
However, we have found that some of the simplest, small $D$, $E$ PEPS wave functions are purely real 
as seen above (hence showing a perfectly momentum-symmetric entanglement spectrum),  
probably due to the fact that they can be re-casted in the form of real fPEPS wave functions,
as it is the case for the fRVB state. In this respect, we found that the ES 
of  the NN fRVB state is completely degenerate, hence saturating the upper bound of the entanglement
measures (e.g., the entanglement entropy per unit length $-{\rm Tr} (\rho\ln{\rho})/N_v$).

\begin{figure}
\begin{center}
\includegraphics[width=\columnwidth,angle=0]{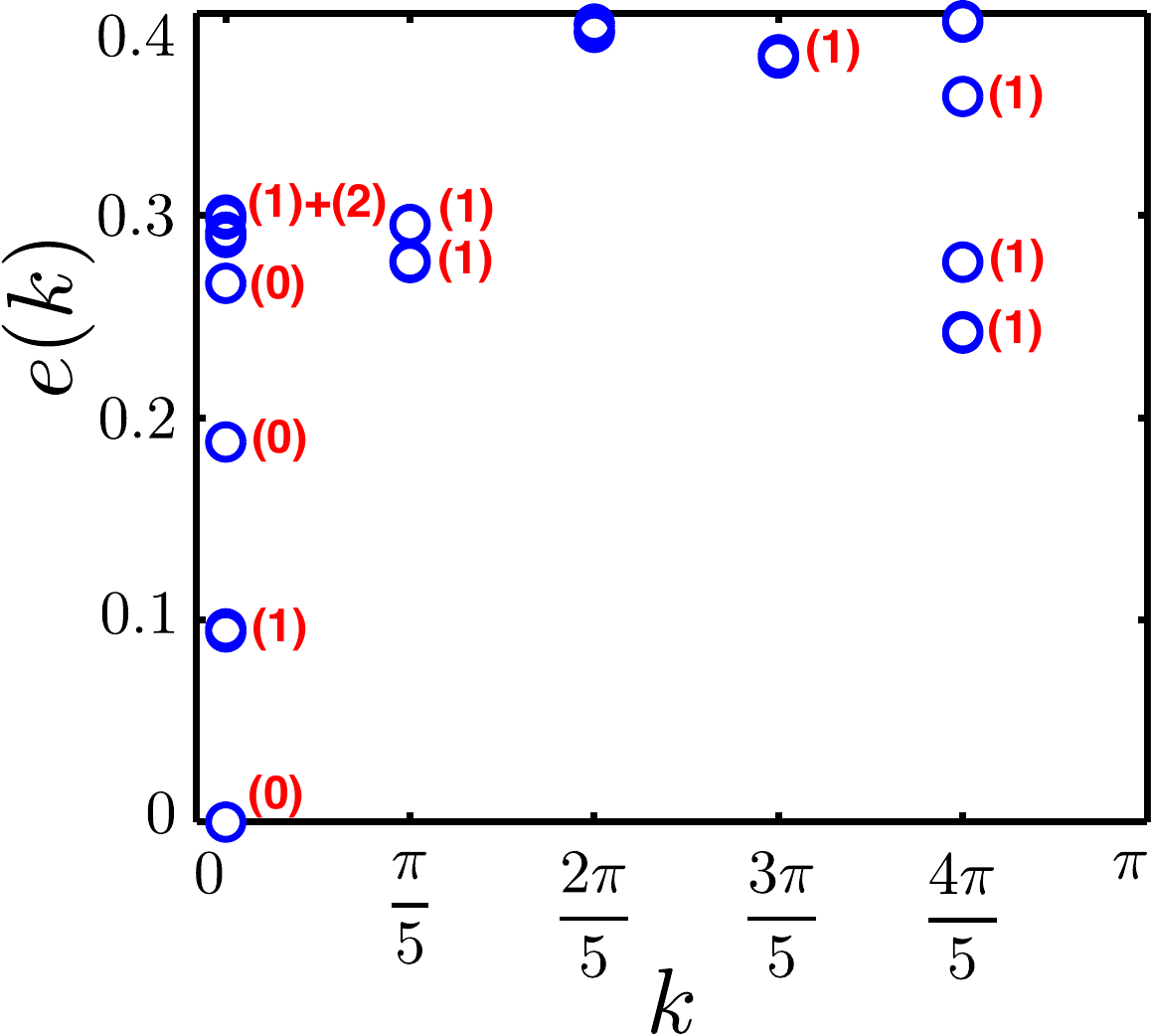}
\caption{[Color online] Entanglement spectrum computed on a $N_v=10$ (infinite) cylinder (with $\chi=20$) 
of the PEPS constructed from the $S=1$, $V=1$, $E^{(a)}$ tensor. Data is plotted vs momentum (modulus $\pi$) along cylinder circumference. Note the spectrum is not symmetric w.r.t. $k=\pi/2$, reflecting TR symmetry-breaking of the PEPS. States can be grouped into SU(2)-multiplets, and $(s)$ labels $2s+1$ degenerate states. 
Note, the spectrum does not seem to fit a simple (chiral) CFT like SU(2)$_2$. }
\label{Fig:ES_E}
\end{center}
\end{figure}

In addition, all {\it complex} $E$ PEPS we have tested turned out to show no well-defined chiral edge modes, although breaking TR symmetry.
For example, this is the case of the PEPS associated to the $S=1$, $V=1$, $E^{(a)}$ tensor
(see Table~\ref{Table:tensor_list} and Supplemental Material)
whose ES is shown in Fig.~\ref{Fig:ES_E}. 
In fact, it is not clear however whether any $E$ tensor could give rise to a topological CSL whenever $S>1/2$. 

\subsection{The $A_1$ + i $A_2$ PEPS} 

Inspired by previous studies, another promising route to construct CSL is to consider tensors of mixed $A_1 + i A_2$ (point group) symmetry,
where both the real and imaginary components can be a sum of tensors belonging to any given $A_1$ and $A_2$ classes involving
identical virtual and physical degrees of freedom. This is a direct generalization of Eq.~(\ref{Eq:chiral_sl}).
This procedure guaranties that the PEPS -- if complex -- transforms into its (orthogonal) complex conjugate state under
any of the reflection symmetries of Fig.~\ref{FIG:lattice_tensor}(a), a necessary condition for a CSL. 
However, this construction does not necessarily imply that the PEPS breaks TR symmetry and the wave function
can, in some cases, be purely real (up to a global phase). In addition,
even if the PEPS breaks TR symmetry, there is no guarantee that it exhibits 
linear dispersing edge modes described by a CFT characteristic  of a topological CSL.
Before describing a successful case (the double-layer CSL and PEPS connected to it), we will show 
below an example of some failure. 

\subsection{Naive spin-$S$ generalizations of the spin-$\frac{1}{2}$ CSL}

It is known that some families of critical (non-chiral) spin-$S$ chains~\cite{Takhtajan1982,Babujian1982} bears a low-energy description in terms of the SU(2)$_k$ Wess--Zumino--Witten models with levels $k=2S$. It is therefore tempting to speculate that
(chiral) SU(2)$_k$ edge modes can originate from $V_i=\frac{k}{2}$ virtual spins
effectively interacting on the edge. Indeed, the $S=\frac{1}{2}$ CSL discussed above, which bears SU(2)$_1$ edge modes, 
involves $V=\frac{1}{2}\oplus 0$ virtual states. In fact, inspecting Table~\ref{Table:tensor_list}, one sees that the 
construction for $S=\frac{1}{2}$ can be easily extended for any spin-$S$ assuming $V=S\oplus 0$~:
one can always combine the unique $n_{\rm occ}=\{1,3\}$ spin-$S$ RVB tensor (in bold case in Table~\ref{Table:tensor_list}) with both the $n_{\rm occ}=\{3,1\}$ $A_1$ tensors (with a real coefficient)
and the $n_{\rm occ}=\{3,1\}$ $A_2$ tensor(s) 
(with a pure-imaginary coefficients).  We have computed the ES at a few points of the 3-dimensional (4-dimensional)
family of PEPS corresponding to the case $S=1$ ($S=\frac{3}{2}$) which turned out to be always gapped, ruling out chiral topological order. 
 
\begin{figure}
\begin{center}
\includegraphics[width=\columnwidth,angle=0]{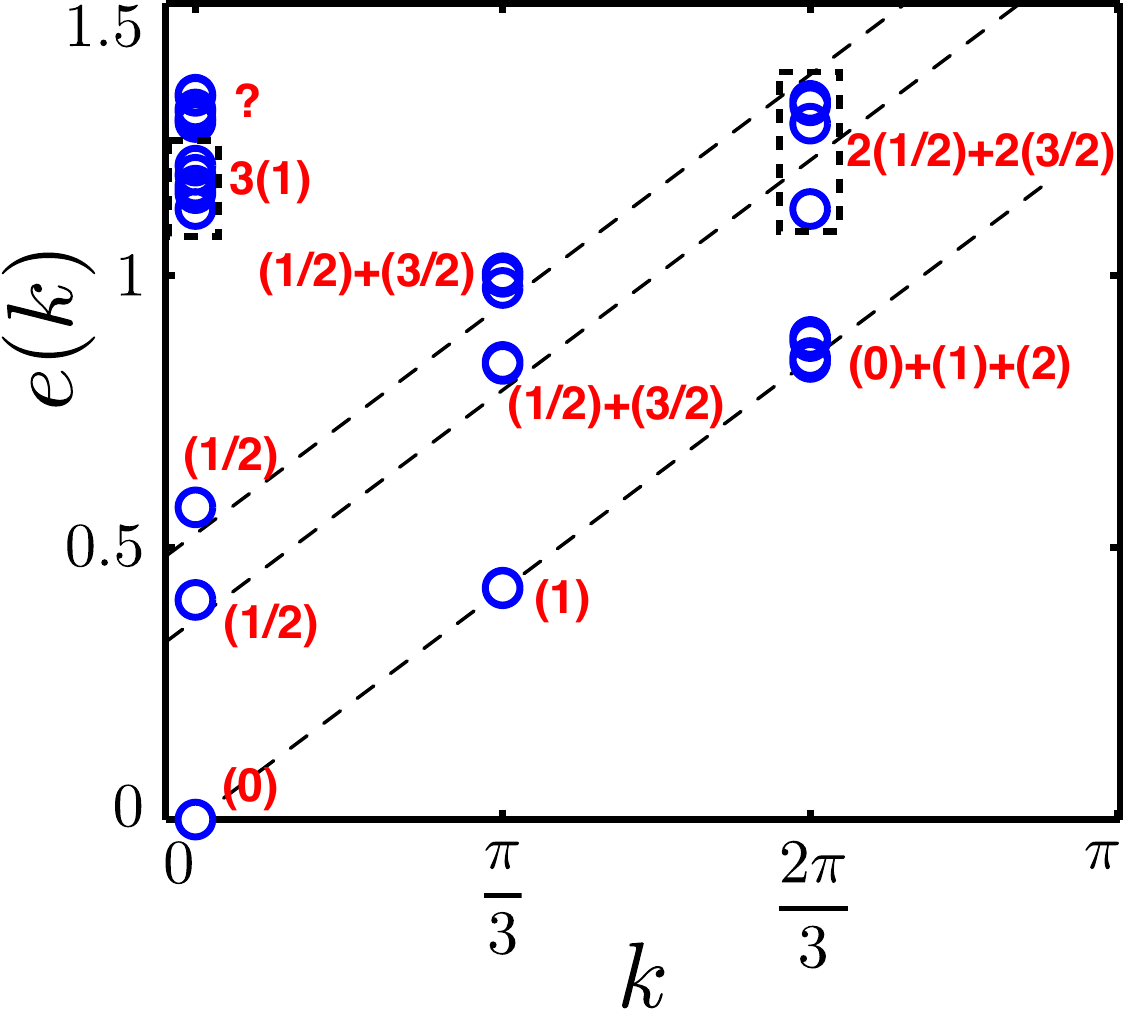}
\caption{Same as Fig.~\protect\ref{Fig:ES_E} for the PEPS obtained from the double-layer tensor of Eq.~\protect\ref{Eq:double_tensor}, 
$N_v=6$ and $\chi=14$. Whenever possible, the multiplet content of the levels is shown, following notations of Table~\ref{Table:su2_2}. The dashed lines are guides to the eye emphasizing the linear dispersion of the modes. }
\label{Fig:ES_double}
\end{center}
\end{figure}

\begin{table}[htb]
\begin{center}
\resizebox{0.95\columnwidth}{!}{%
  \begin{tabular}{@{} cccc @{}}
    \hline \hline
    $n \backslash j $ & 0 & $\frac{1}{2}$ & 1 \\ 
  \hline 
  & & &\\
0 & (0) {\color{red} [1]}& ($\frac{1}{2}$) {\color{red} [2]} &(1) {\color{red} [3]}
\\
1 & (1) {\color{red} [3]} & ($\frac{1}{2}$)+($\frac{3}{2}$) {\color{red} [6] } & (0)+(1) {\color{red} [4]}
\\
2 & (0)+(1)+(2) {\color{red} [9]} & 2($\frac{1}{2}$)+2($\frac{3}{2}$) {\color{red} [12]} & (0)+2(1)+(2) {\color{red} [12]}
\\
3 & (0)+3(1)+(2) {\color{red} [15]} & 4($\frac{1}{2}$)+3($\frac{3}{2}$)+($\frac{5}{2}$) {\color{red} [26]} & 2(0)+3(1)+2(2) {\color{red} [21]}
\\
& & &\\
    \hline 
    \hline
      \end{tabular}
     }
\caption{[Color online] Towers of states of the SU(2)$_2$ WZW model, in each of the three sectors characterized by
the primary fields $j=0,\frac{1}{2},1$ (listed in each column) and conformal weights $\frac{1}{4}j(j+1)$. Each line corresponds to a Virasoro 
level indexed by $n$. For each sector and each level, the (quasi-) degenerate states can be grouped in terms of exact SU(2) multiplets
like $n_0 (0) + n_1 (1) +\cdots$ (meaning $n_0$ singlets, $n_1$ triplets, etc...). The (red) numbers in brackets 
correspond to the total number of states in each group of levels. 
}
\label{Table:su2_2}
\end{center}
\end{table}

\subsection{Double-layer CSL and SU(2)$_2$ edge states}

\begin{figure}
\begin{center}
\includegraphics[width=0.65\columnwidth,angle=0]{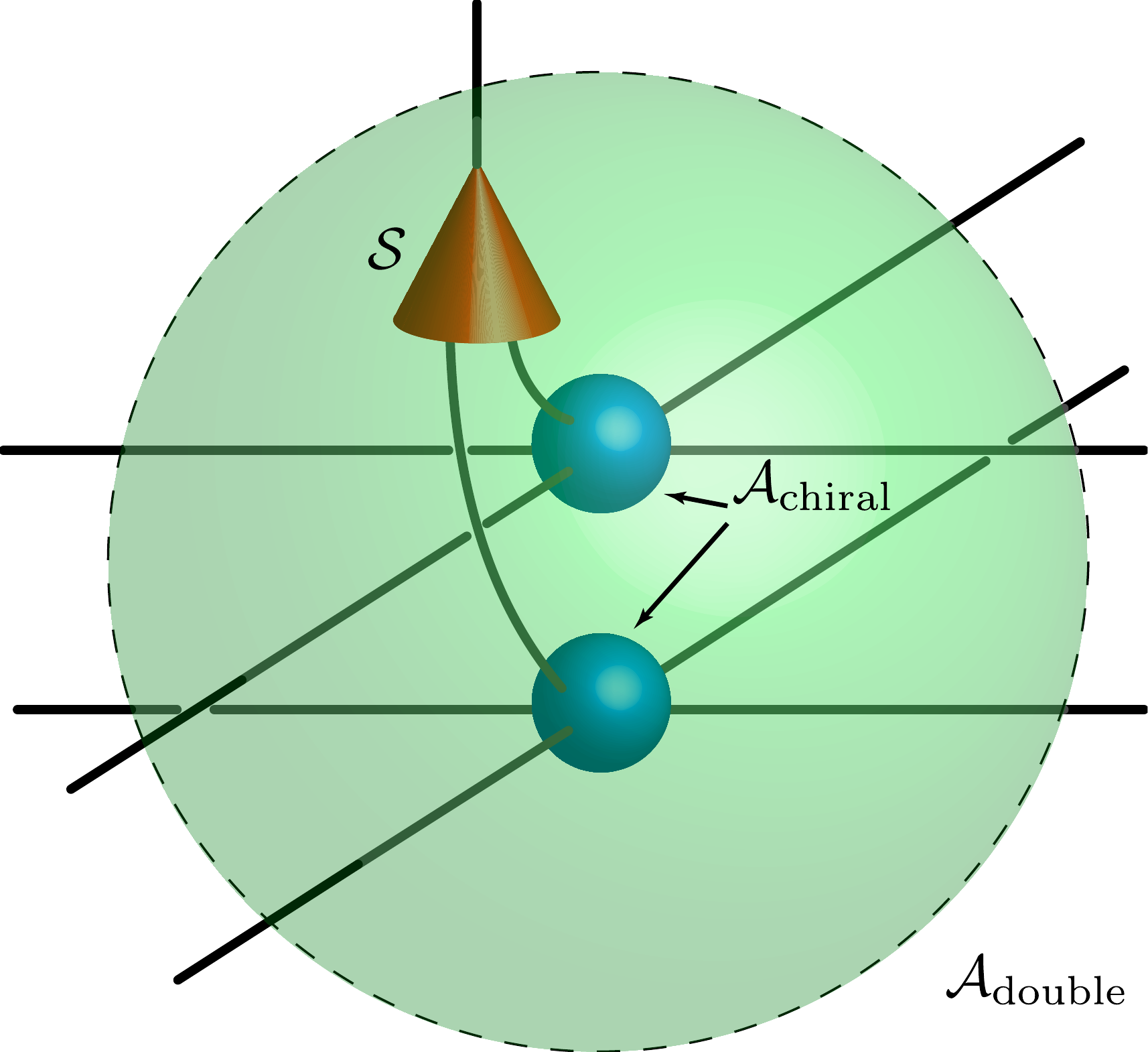}
\caption{[Color online] Diagram corresponding to Eq.~\ref{Eq:double_tensor}~: the two layers of tensors ${\cal A}_{\rm chiral}$ are symmetrized by the isommetry $S$, which projects the physical dimensions in the 3-dimensional spin-$1$ subspace of $\frac{1}{2} \otimes \frac{1}{2}$, giving rise to the tensor ${\cal A}_{\rm double}$}
\label{FIG:Doub}
\end{center}
\end{figure}

To go beyond the above naive construction, we shall follow here a strategy borrowed from the field of the 
Fractional Quantum Hall States (FQHS)~\cite{Laughlin1983,Moore1991}.
Recently, using interpretation of FQHS as conformal blocks in certain rational conformal field theories~\cite{Moore1991}, MPS representations of FQHS were exploited~\cite{Estienne2013a,Estienne2013b}, providing unprecedented 
numerical accuracy~\cite{Estienne2015}. 
While the (Abelian) Laughlin state can be written as a simple MPS, the non-Abelian states can be constructed as multilayer fractional quantum Hall wave functions upon symmetrization over the layer index~\cite{Greiter2009,Repellin2015}. 
Very recently, symmetrization of topologically ordered PEPS was shown to be a powerful method for constructing new
topological models~\cite{Gonzales2016}. Here, we will apply this procedure 
by considering {\it two layers} of the CSL defined by Eq.~(\ref{Eq:chiral_sl}).
The double-layer tensor $[{\cal A}_{\rm chiral}]^{\otimes 2}$
is symmetrized w.r.t the $\frac{1}{2}\otimes\frac{1}{2}$ physical variables, 
\begin{equation}
{\cal A}_{\rm double}={\cal S}[{\cal A}_{\rm chiral}\otimes {\cal A}_{\rm chiral}] \, ,
\label{Eq:double_tensor}
\end{equation}
hence projecting onto the $S=1$ physical state (see Fig.~\ref{FIG:Doub}). 
More precisely, its $S_z=+1,0,-1$ (physical) components are given by,
\begin{eqnarray}
{\cal A}_{\rm double}\vert_1&={\cal A}_{\rm chiral}\vert_{\frac{1}{2}}\otimes {\cal A}_{\rm chiral}\vert_{\frac{1}{2}} , \nonumber \\
{\cal A}_{\rm double}\vert_0&=
\frac{1}{\sqrt{2}} \{{\cal A}_{\rm chiral}\vert_{\frac{1}{2}}\otimes {\cal A}_{\rm chiral}\vert_{-\frac{1}{2}} \nonumber \\
&+ {\cal A}_{\rm chiral}\vert_{-\frac{1}{2}}\otimes {\cal A}_{\rm chiral}\vert_{+\frac{1}{2}}\},  \\
{\cal A}_{\rm double}\vert_{-1}&={\cal A}_{\rm chiral}\vert_{-\frac{1}{2}}\otimes {\cal A}_{\rm chiral}\vert_{-\frac{1}{2}} , \nonumber 
\end{eqnarray}
where the virtual variables of the double-layer tensor on the left hand side are given by the tensor product of the virtual variables of the
two single layer CSL tensors on the right hand side. For convenience, we then realize a (unitary) change of the $D=9$ virtual basis, 
from the $[\frac{1}{2}\oplus 0]_{\rm top}\otimes[\frac{1}{2}\oplus 0]_{\rm bottom}$ natural basis to the ``symmetric" basis 
$1\oplus\frac{1}{2}\oplus\frac{1}{2}\oplus 0_s\oplus 0$ described in Fig.~\ref{FIG:single_double}(b),
where the two spin-$\frac{1}{2}$-representations correspond now to spin-$\frac{1}{2}$ states symmetric and antisymmetric w.r.t. layer exchange, respectively, and the spin-$0_s$ and spin-$0$ representations contain the $\frac{1}{\sqrt{2}}(\uparrow\downarrow-\downarrow\uparrow)$ and  $00$ singlets, respectively.
It can be seen easily (see later for details) that ${\cal A}_{\rm double}$ inherits from ${\cal A}_{\rm chiral}$
SU(2)-spin rotation symmetry and lattice $A_1+iA_2$ (orbital) symmetry. It is therefore expected to break TR symmetry while preserving all
lattice symmetries, a key property of chiral spin liquids. 

We have computed the ES of the double-layer tensor for $\lambda_1=\lambda_2=\lambda_c(=1/\sqrt{3})$ on an infinite $N_v=6$ cylinder and results are shown in Fig.~\ref{Fig:ES_double}.
Linearly dispersing branches are clearly seen. Lot of resemblance with the chiral SU(2)$_2$ CFT spectrum is seen. The later shown in Table~\ref{Table:su2_2} contains 
three sectors labeled by the primary fields $j=0,\frac{1}{2},1$. The lowest branch of the ES agrees perfectly with the
content of the $j=0$ sector. We also observe two almost degenerate branches with some energy offset, but with the same slope,
each compatible with the theoretical expectation for the 
$j=1/2$ sector.  It should be noted that, although an energy offset of $3/16$ of the (average) level spacing 
is expected for the $j=1/2$ branch, a larger offset (by a factor $\sim 4$) is observed which could be plausibly attributed to finite 
perimeter ($N_v$) and finite-$\chi$ effects in the ES calculation (see Appendix \ref{app1}). 

\begin{figure}[htbp]
\begin{center}
\includegraphics[width=0.9\columnwidth,angle=0]{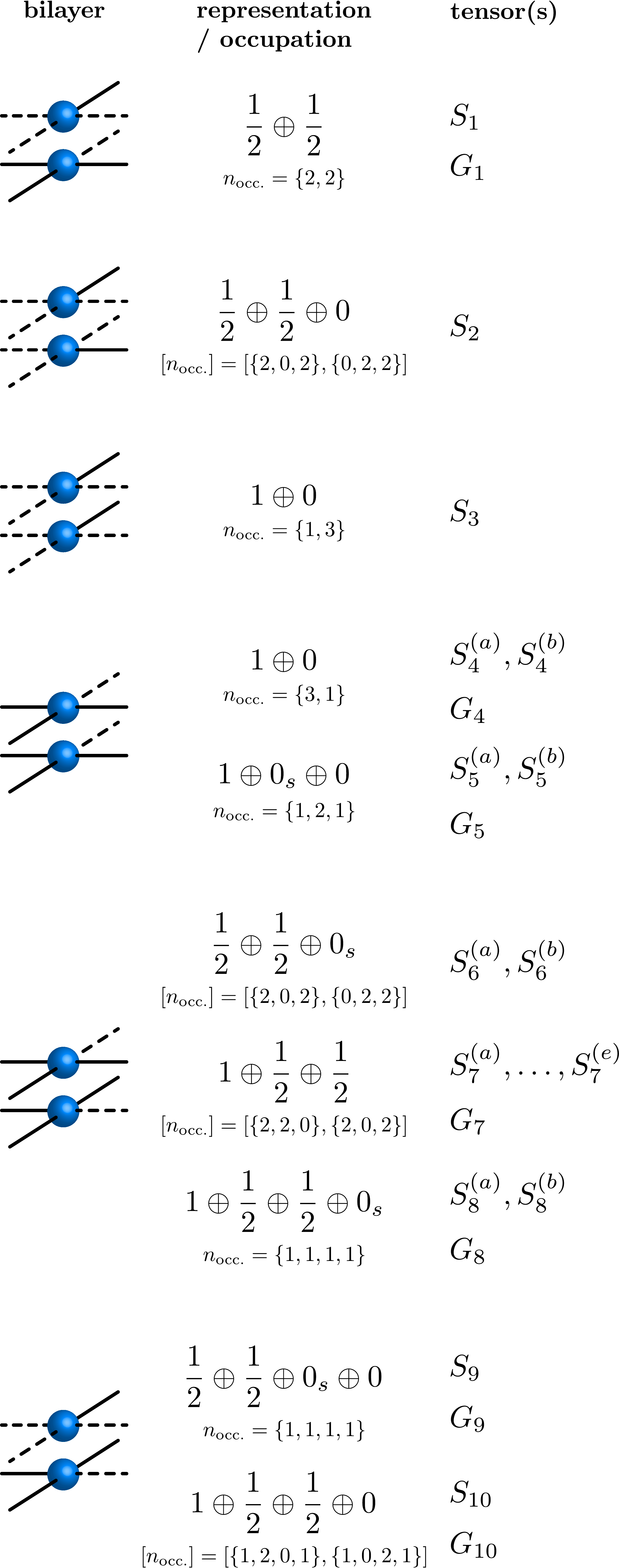}
\caption{
All combinations (up to global $\pm\pi/2$ or $\pi$-rotations) obtained by expending the ${\cal A}_{\rm chiral}$
tensors (according to Eq.~(\ref{Eq:chiral_sl}))
simultaneously in the top and bottom layers of the (symmetrized) double-layer CSL (left-most column),
leading to
orthonormal tensors (right-most column) belonging to different representations of the virtual states 
(middle column). 
 }
\label{Fig:double_layer}
\end{center}
\end{figure}

\subsection{Decomposition in terms of elementary tensors}

The double-layer tensor involves $D=9$ virtual states, making hard the computation of the ES on larger cylinders and with
larger MPS dimension $\chi$ to permit a more definite assignment of the edge theory. In addition, it is not clear whether the 
observation of exact SU(2)$_2$ edge modes requires some degree of ``fine tuning".
For these two reasons, it is a good idea to try to construct simpler (i.e., with lower bond dimension $D$) PEPS 
which, potentially, could exhibit chiral edge modes.
Our strategy is here to ``break up'' the double-layer tensor ${\cal A}_{\rm double}$ into independent parts 
defined by smaller virtual spaces $V$
but still exhibiting SU(2)-spin rotation and $A_1 + i A_2$-lattice symmetries. To do so, we expend the ${\cal A}_{\rm chiral}$
tensors according to Eq.~(\ref{Eq:chiral_sl}) simultaneously in the top and bottom layers.
Fig.~\ref{Fig:double_layer} shows all possible combinations depending on the relative orientation of the various contributions
in the two layers. In fact, it can be shown that each part leads to the sum of a few orthonormal SU(2)-spin symmetric real 
$S$ and $G$ tensors, whose virtual states belong to a smaller representation, subset of the overall Hilbert space  
$V=[\frac{1}{2}\oplus 0]^{\otimes 2}$.
The $S$-tensors and $G$-tensors belong to the $A_1$ (s-wave) and $A_2$ (g-wave) IRREP and appear with real and pure-imaginary coefficients,
respectively, so that each part of the decomposition
bears an overall $A_1+i A_2$ symmetry. More precisely, we can write
\begin{equation}
{\cal A}_{\rm double} = \sum_{\alpha} s_\alpha S_\alpha + \sum_{\beta} g_\beta G_\beta
\label{Eq:decomposition}
\end{equation}
with
\begin{align} 
s_1 & =  \frac{\lambda _1 \lambda _2}{\sqrt{2}} &  
s_2 & =  \frac{1}{2} \sqrt{3} \lambda _1^2 \nonumber\\
s_3 & =  \frac{\lambda_1^2}{2} &
s_4^{(a)} &=  \frac{7 \lambda _2^2-3 \lambda _c^2}{12 \sqrt{2}} \nonumber\\
s_4^{(b)} &=  \frac{1}{12} \sqrt{\frac{5}{2}} \left(\lambda _2^2+3 \lambda _c^2\right) &
s_5^{(a)} &= \frac{\lambda _2^2}{3 \sqrt{2}} \nonumber\\
s_5^{(b)} &=  \frac{1}{12} \left(-\lambda _2^2-3 \lambda _c^2\right)  &
s_6^{(a)} &=  \frac{5 \lambda _2^2+3 \lambda_c^2}{12 \sqrt{2}} \nonumber\\
s_6^{(b)} &=  \frac{1}{12} \left(\lambda _2^2-3 \lambda _c^2\right)&
s_7^{(a)} &= -\frac{\lambda _2^2+9 \lambda_c^2}{12 \sqrt{3}} \nonumber\\
s_7^{(b)} &=  \frac{1}{3} \sqrt{\frac{5}{3}} \lambda _2^2 &
s_7^{(c)} &=  \frac{17 \lambda _2^2+15 \lambda _c^2}{36 \sqrt{2}}\nonumber\\
s_7^{(d)} &=  -\frac{25 \lambda _2^2+3 \lambda_c^2}{18 \sqrt{13}} & 
s_7^{(e)} &=  -\frac{1}{3} \sqrt{\frac{5}{26}} \left(\lambda _2^2-3 \lambda _c^2\right) \nonumber\\
s_8^{(a)} &=  \frac{9 \lambda _2^2-5 \lambda _c^2}{12 \sqrt{3}} &
s_8^{(b)} &=  \frac{1}{3} \sqrt{\frac{7}{6}} \lambda _c^2 S_8^{(b)} \nonumber\\
s_9 & =  \frac{\lambda _1 \lambda _2}{2} &
s_{10} & =  \frac{1}{2} \sqrt{5} \lambda _1 \lambda _2 \nonumber\\
g_1 & = - \frac{i \lambda _1 \lambda _c}{\sqrt{2}} &
g_4 & = -\frac{i \lambda_2 \lambda _c}{\sqrt{3}} \nonumber\\
g_5 & =  -\frac{i \lambda _2 \lambda_c}{\sqrt{6}} &
g_7 & =  - i \sqrt{\frac{5}{6}} \lambda _2 \lambda _c\nonumber\\
g_8 & =  i \sqrt{\frac{2}{3}} \lambda_2 \lambda _c &
g_9 & =  -\frac{1}{2} i \lambda _1 \lambda _c \nonumber\\
g_{10} & = \frac{1}{2} i \sqrt{5} \lambda _1 \lambda _c &&\nonumber
\label{Eq:decomposition_coeff}
\end{align}
where the subscripts of the $S$ ($s$) and $G$ ($g$) tensors (coefficients) label the different virtual spin representations according 
to Fig.~\ref{Fig:double_layer}. The exact expressions of all $S$ and $G$ spin-$1$ tensors are providing in Appendix B, written in the same
$D=9$ overall basis so that any linear combination of tensors can easily be performed. 

\begin{figure*}
\begin{center}
\includegraphics[width=1.8\columnwidth,angle=0]{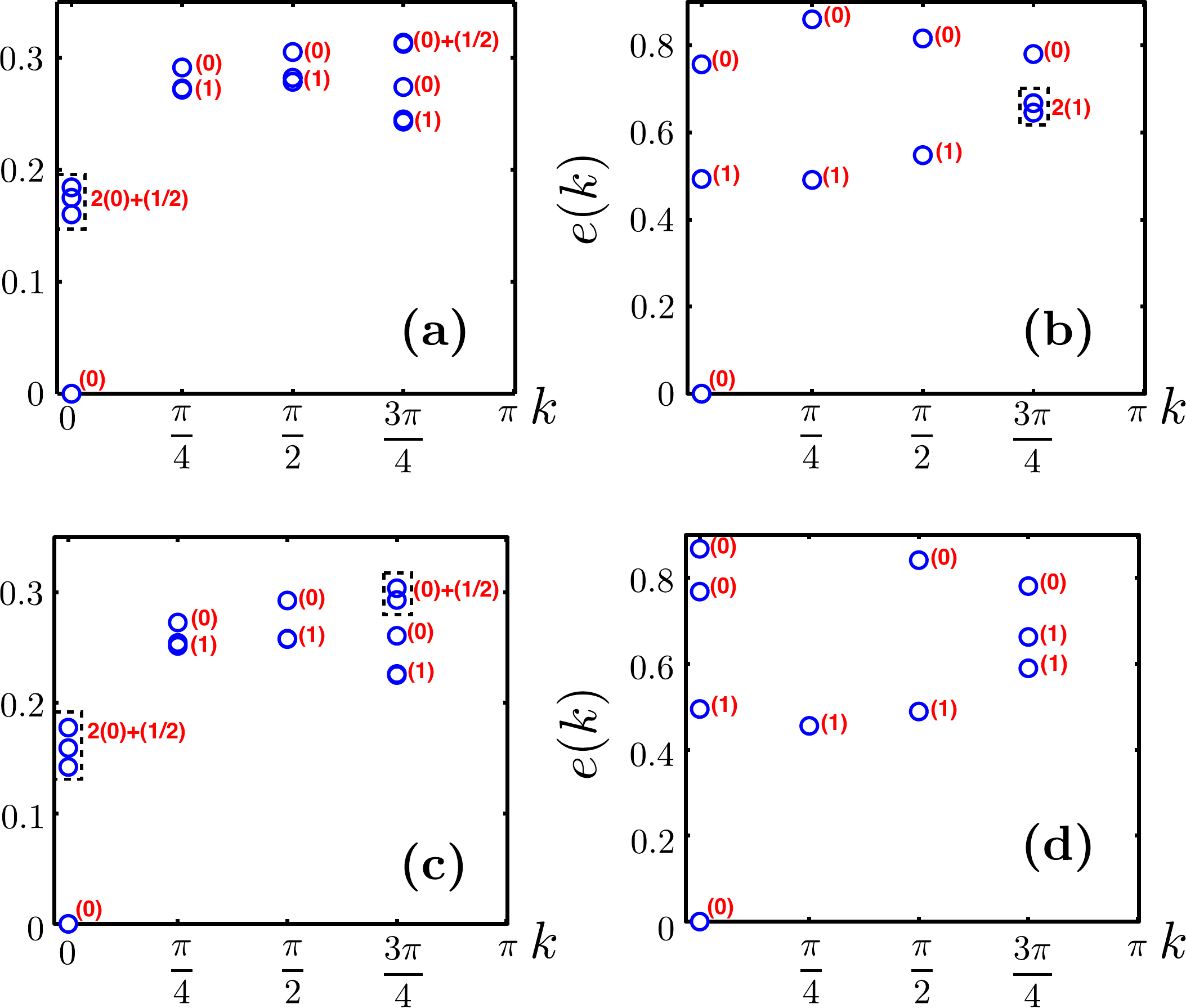}
\caption{[Color online] ES for the PEPS obtained from various subsets of Eq.~(\protect\ref{Eq:decomposition}), and $N_v=8$, $\chi=20$~:
(a) $S_1$ and $G_1$ tensors ($\frac{1}{2}\oplus\frac{1}{2}$, $D=4$);
 (b) $S_3$, $S_4$ and $G_4$ tensors ($1\oplus 0$, $D=4$);
 (c) $S_1$, $S_6$ and $G_1$ tensors ($\frac{1}{2}\oplus\frac{1}{2}\oplus 0_s$, $D=5$);
 (d) $S_3$, $S_4$, $S_5$, $G_4$ and $G_5$ tensors ($1\oplus 0\oplus 0_s$, $D=5$).}
\label{Fig:ES_partial}
\end{center}
\end{figure*}

\begin{figure*}[htbp]
\begin{center}
\includegraphics[width=1.8\columnwidth,angle=0]{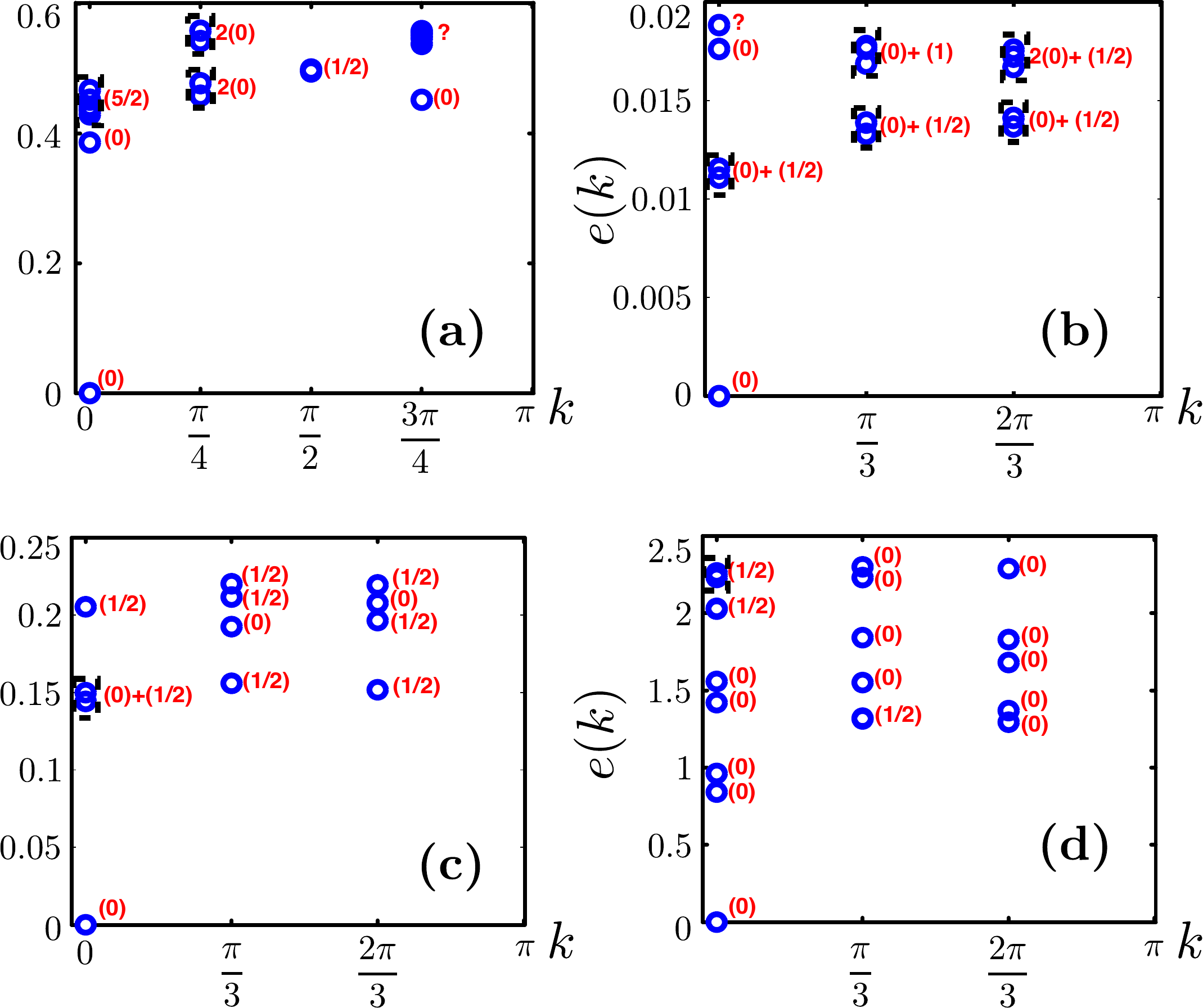}
\caption{Same as Fig.~\protect\ref{Fig:ES_partial} for other subsets of 
Eq.~(\protect\ref{Eq:decomposition}), and $\chi=10$~:
 (a) $N_v=8$, for $S_1$, $G_1$, $S_2$, $S_6$, $S_9$ and $G_9$
tensors ($\frac{1}{2}\oplus\frac{1}{2}\oplus 0\oplus 0_s$, $D=6$);
(b) $N_v=6$, for $S_1$, $G_1$, $S_7$ and $G_7$ tensors ($1\oplus\frac{1}{2}\oplus\frac{1}{2}$, $D=7$);
(c)  $N_v=6$, for $S_1$, $S_6$, $S_7$, $G_1$, and $G_7$ ($1\oplus\frac{1}{2}\oplus\frac{1}{2}\oplus 0_s$, $D=8$);
(d)  $N_v=6$, for $S_1$, $S_2$, $S_3$, $S_4$, $S_7$, $S_{10}$, $G_1$, $G_4$, and $G_7$ ($1\oplus\frac{1}{2}\oplus\frac{1}{2}\oplus 0$, $D=8$).
}
\label{Fig:ES_partial2}
\end{center}
\end{figure*}

We have applied the above decomposition (\ref{Eq:decomposition}) for the same choice of the parameters  $\lambda_1=\lambda_2=\lambda_c(=1/\sqrt{3})$ and computed the ES of a few ad-hoc linear combinations of $S$ and $G$ tensors (to keep the $A_1+iA_2$
symmetry) of larger and larger bond dimension $D$. 
Keeping only the $S_1$ and $G_1$ tensors, on one hand, or the $S_3$, $S_4$ and $G_4$ tensors on the other hand,
enables to restrict the bond dimension to the same $D=4$ small value, although the two PEPS involve completely different 
virtual degrees of freedom, $\frac{1}{2}\oplus\frac{1}{2}$ in the first case and $1\oplus 0$ in the second case. 
We have found that their corresponding ES shown in Fig.~\ref{Fig:ES_partial} (a) and Fig.~\ref{Fig:ES_partial}(b), respectively, seem 
to be both gapped. 

By adding, gradually, more virtual degrees of freedom to the previous cases, it is interesting to see whether the gap
in the ES closes. For instance, adding the $S_6$ tensors to the $S_1$ and $G_1$ tensors enlarge the 
virtual space to $\frac{1}{2}\oplus\frac{1}{2}\oplus 0_s$ and $D=5$. However, as seen in Fig.~\ref{Fig:ES_partial}(c), the gap seems to persist.
Similarly, adding the $S_5$ and $G_5$ tensors to the $S_3$, $S_4$ and $G_4$ tensors enlarge the 
virtual space to $1\oplus 0\oplus 0_s$ and $D=5$, but does not close the gap either (see Fig.~\ref{Fig:ES_partial}(d)). 

Next, we have considered the virtual representations $V=\frac{1}{2}\oplus\frac{1}{2}\oplus 0\oplus 0_s$ ($D=6$)
involving the $S_1$, $G_1$, $S_2$, $S_6$, $S_9$ and $G_9$ tensors,
$V=1\oplus\frac{1}{2}\oplus\frac{1}{2}$ ($D=7$) 
involving the $S_1$, $G_1$, $S_7$ and $G_7$ tensors,  $V=1\oplus\frac{1}{2}\oplus\frac{1}{2}\oplus 0_s$ ($D=8$)
involving the $S_1$, $S_6$, $S_7$, $G_1$, and $G_7$ tensors and 
$V=1\oplus\frac{1}{2}\oplus\frac{1}{2}\oplus 0$ ($D=8$) involving the $S_1$, $S_2$, $S_3$, $S_4$, $S_7$, $G_1$, $G_4$, and $G_7$ tensors.  Results are compared in Fig.~\ref{Fig:ES_partial2}(a-d). 
As the number of tensors and the bond dimension increase, one does not observe any systematic trend, rather 
the ES changes in some erratic fashion. The later seems nevertheless to remain gapped,
although in Fig.~~\ref{Fig:ES_partial2}(b) the (pseudo-)energy scale becomes very small.
In any case, the ES remains very different from the chiral ES of the $D=9$ double-layer CSL 
shown in Fig.~\ref{Fig:ES_double}. This is a clear indication that there is a large degree of fine tuning in the latter wave function.
In other words, just imposing SU(2)-spin rotation and $A_1+iA_2$ orbital symmetries, as it was done for 
the single layer CSL, is not sufficient to obtain a CSL. One reason might be that for SU(2)$_2$ CFT, 
(i) spin-1 degrees of freedom would be needed on the boundary but 
(ii) spin-1 chains are generically in the gapped Haldane phase.
A gapless spectrum in a spin-1 chain requires fine tuning~\cite{Takhtajan1982,Babujian1982}.

\section{Conclusions and outlook}
\label{sec4} 

In this paper we have elaborated a classification scheme of all rank-5 SU(2)-symmetric tensors according to the on-site physical spin $S$, the local Hilbert space of the bond degrees of freedom, and the irreducible representations of the $C_{4v}$ point group of the square lattice. We have shown how many remarkable (Mott insulating) states of matter fall naturally into this classification. More generally, we have explained how our scheme can be used to systematically construct 
families of translationally invariant many-body singlet states, preserving or breaking discrete (point group)
lattice symmetries, spin liquids and (lattice) nematics, respectively. However, we bring here a few words of caution~: first, we should mention that LRO
(associated e.g. to spontaneous translation and/or SU(2) symmetry breaking) may still appear 
in the thermodynamic limit in some parameter regions, 
the PEPS being in that case a fully symmetric ``Shr\"odinger cat state''. 
Note that the existence of LRO in our symmetric PEPS can only be diagnosed by a 
thorough numerical investigation, e.g. inspecting the low-energy spectrum 
of the transfer operator~\cite{Schuch2013a}.
Secondly, it is likely that 
not all translationally invariant and spin-rotationally symmetric spin liquids (on the square lattice) 
can be expressed in terms of a PEPS based on a single on-site SU(2)-symmetric tensor. 
However, we believe our classification encompasses
a very large manifold of symmetric spin liquids. Spin liquids not generated by our classification
may include e.g. those requiring a two-site (gauge) unit cell such as the (translationally invariant) $\pi$-flux PEPS 
of the PSG classification~\cite{Jiang2015} or those requiring a different type of virtual particles like
fermions, Majoranas, anyons, etc...

We have also used our construction to systematically search for higher-spin ($S>1/2$) topological chiral spin liquids. One of our constructions uses a symmetrization over a double-layer PEPS, showing gapless chiral edge modes corresponding to a non-Abelian SU(2)$_2$ Wess-Zumino-Witten model, which we have 
determined via the analysis of its entanglement spectrum.  
This family of CSL can be seen as a 2-dimensional manifold (spanned by the parameters $\lambda_2/\lambda_1$
and $\lambda_c/\lambda_1$) imbedded in a much larger PEPS family (characterized by arbitrary superpositions of 
the $S$ and $G$ tensors). This suggests that, more generally, non-Abelian CSL live on (relatively small) 
fine-tuned manifolds of large PEPS families. Also, we believe that our construction can be extended to $k$-layers, 
then showing SU(2)$_k$ gapless chiral edge modes.

We envisage these results as the first step of a broader research program, concerning the search of quantum spin liquids with tensor networks. Further work will involve, for instance, using the tensors produced in our classification in order to produce \emph{ans\"atze} for the numerical simulation of the frustrated Heisenberg model on the square lattice (improving the results of Ref.~\cite{Wang2013}), extending the classification scheme to the Kagome lattice, and using it to propose variational wave functions to search for new spin liquids for the Kagome Heisenberg Antiferromagnet. 

Another straightforward extension of our classification is to enlarge the physical space to more 
degrees of freedom e.g. to include extra U(1)-charge degrees of freedom~\cite{Poilblanc2014} to describe hole 
doping a Mott (spin liquid or antiferromagnetic) insulator, or more orbital degrees of freedom ($N>2$ with SU(N) symmetry).
To build wavefunctions with  specified U(1)-charge, covariant tensors can be constructed.
Lastly, we note that our PEPS are, by construction, translationally invariant. However, it is also straightforward to
describe states breaking translation invariance by considering different tensors at every site of a given supercell.
All these problems, and more, will be addressed in future work along these lines. 

\bigskip 

\begin{acknowledgments}

R.O. acknowledges the Laboratoire de Physique Th\'eorique, C.N.R.S. and Universit\'e de Toulouse, for hosting him during the period in which this work was initiated, as well as the MOGON computer cluster of the University of Mainz for computational resources. D.P. is supported by the TNSTRONG ANR (French Research Council) grant (2016-2020).
D.P. thanks Nicolas Regnault and German Sierra for help in completing Table~\ref{Table:su2_2}.  

\end{acknowledgments}

\bibliography{bibliography}

\clearpage
\appendix

\section{Entanglement Spectrum from 2d PEPS}
\label{app1}

2d PEPS are a natural arena to study the so-called Entanglement Spectrum (ES)~\cite{Li2008}, 
see for instance Refs.~\cite{Cirac2011, Yang2014}. In this appendix we review briefly what are the main ingredients of the calculations of ES in the context of 2d PEPS with tensor network methods, which we have used in several sections of this paper to characterize the edge modes.  

\begin{figure}
\begin{center}
\includegraphics[width=1.007\columnwidth,angle=0]{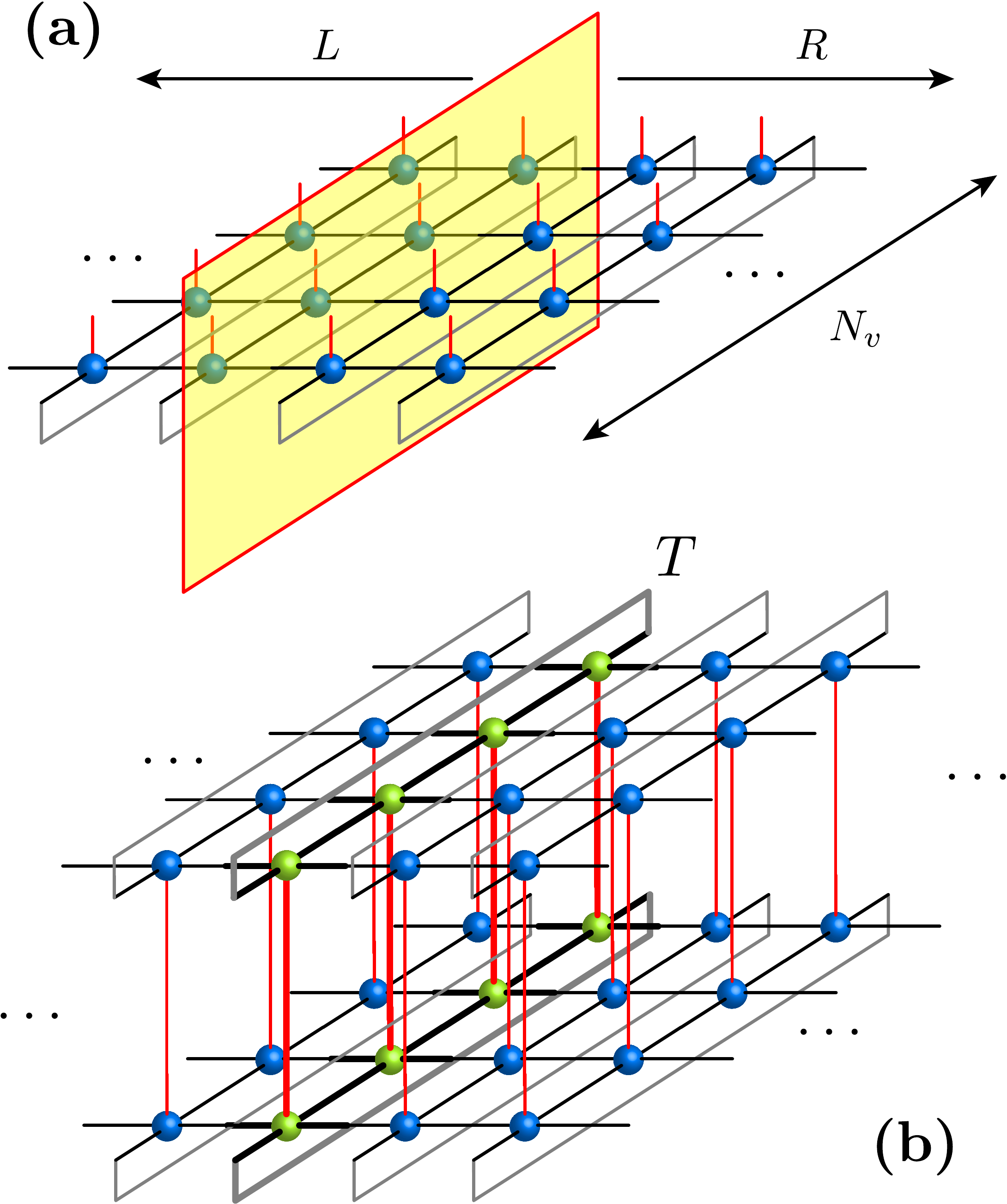}
\caption{[Color online] (a) 2d PEPS wrapped around a cylinder of width $N_v$, and splitted into two parts $L$ and $R$. (b) Transfer matrix $T$ of the PEPS.}
\label{Fig:App}
\end{center}
\end{figure}

Consider a 2d PEPS $\ket{\Psi}$ wrapped around a cylinder of circumference $N_v$, as in Fig.~\ref{Fig:App}(a). For the sake of this calculation, we consider the cylinder to be infinitely-long \cite{Cirac2011}. We now split the cylinder in two parts, say, left ($L$) and right ($R$). As explained in Ref.~\cite{Cirac2011}, the reduced density matrix of half an infinite cylinder, e.g., for $L$ is given by
\beq
\rho = U \sqrt{\sigma_L^T}\sigma_R \sqrt{\sigma_L^T} U^\dagger,
\label{Eq:rho}
\eeq
with $\sigma_{L/R}$ the reduced density operators in $L/R$ for the virtual spaces across the bipartition, and $U$ an isometry obtained from the contraction of the PEPS tensors. Technically, $\sigma_{L/R}$ are computed as the dominant left/right eigenvectors of the PEPS transfer matrix $T$, see Fig.\ref{Fig:App}(b). From this equation, it is clear that $\rho$ has the same eigenvalues than $\sqrt{\sigma_L^T}\sigma_R \sqrt{\sigma_L^T}$, since both operators are related only by an isometry (which leaves the spectrum invariant). Moreover, $\sqrt{\sigma_L^T}\sigma_R \sqrt{\sigma_L^T}$ shares the spectrum with $\sigma_L^T \sigma_R$, and shares the same quantum numbers for the eigenvectors \cite{Poilblanc2016}. This is particularly convenient, since square-roots are difficult to implement in the context of tensor network methods. 

The calculation of $\sigma_{L/R}$ on an infinite cylinder for a 2d PEPS is a well-posed tensor network problem that  can be solved using many different strategies. Here we use a similar approach to the one used in Refs.\cite{Poilblanc2016, Orus2014_2}. Long story short: we compute this dominant eigenvector using the iTEBD method for non-unitary evolutions \cite{Vidal2007,OrusVidal2008}. More specifically, for a given set of tensors in the transfer matrix $T$, we compute the tensors for $\sigma_{L/R}$ assuming $N_v \rightarrow \infty$ using iTEBD. The resulting dominant eigenvector can be written as an MPO of bond dimension $\chi$, which is then wrapped around a circle of length $N_v$, and constitutes our approximated $\sigma_{L/R}$. This approach is remarkably efficient, and in practice provides very accurate results, including the ones in this paper. The computational cost of this calculation is $O(\chi^3 D^6 + \chi^2 D^8)$ \cite{Orus2014_2}. Moreover, in this calculation it is usually possible to fix topological and/or parity sectors if needed, by feeding this as an input in the initial vector of iTEBD. 

The diagonalization of  $\sigma_L^T \sigma_R$ then proceeds by using Krylov-subspace methods (e.g., Lanczos). Such methods rely on matrix-vector multiplications which, in our case, can be done very efficiently since the matrix $\sigma_L^T \sigma_R$ has a explicit tensor network structure in terms of the MPOs for $\sigma_L$ and $\sigma_R$.  At this point it is also possible to fix the total $z$-component of the spin, which is a good quantum number in our SU(2)-invariant setting, just by fixing it in the initial vector used in Krylov methods. Finally, momentum in the transverse direction is also a good quantum number, which can be extracted a posteriori by checking how the corresponding eigenvectors transform under the action of the translation operator. 

\clearpage
\section{PEPS tensors of simple states}
\label{app2}

\subsection{Conventions for labeling virtual states}

We list below the tensors of the states described in the text. 
We first provide the conventions used to label the virtual states $|\alpha\big>$. Considering a virtual subspace $V=S_1 \oplus \ldots \oplus S_p$ the labels run from $0$ to $2(S_1+\ldots+S_p)+(p-1)$ by decreasing order of $S_z$ as explicitly described in Table \ref{Table:convention}.
\begin{table}[h]
\begin{center}
\small
$
\begin{array}{lc}\hline\hline
 \text{Virtual state} 	& \text{Tensor index} \\
  (S_z) 					&  \\ \hline
S_1  					& 0 \\
S_1-1    				& 1 \\
\ldots  					& \ldots \\
-S_1  					& 2 S_1 \\
S_2  					& 2 S_1+1 \\
S_2-1  					& 2 S_1+2 \\
 \ldots    				& \ldots \\
-S_p      				& 2 \left(S_1+\ldots S_p\right)+(p-1) \\ \hline\hline
\end{array}$
\caption{Conventions for labeling virtual states for $V=S_1 \oplus \ldots \oplus S_p$.}
\label{Table:convention}
\end{center}
\end{table}

\vspace{-0.2cm}
\subsection{The $S=2$ AKLT state} 
\begin{longtable}{NN}\hline \hline
Sz=+2 & Sz=+1   \\ \hline\endfirsthead
\hline \multicolumn{2}{c}{Table \ref{Table:s2AKLT} (continued)} \\ \hline\endhead 
T_2(0,0,0,0)=1 &	T_1(0,0,0,1)=\frac{1}{2}  \\
	&T_1(0,0,1,0)=\frac{1}{2}	 \\
	&T_1(0,1,0,0)=\frac{1}{2}	 \\
	&T_1(1,0,0,0)=\frac{1}{2}	 \\ \hline
Sz=0  & \\ \hline
T_0(0,0,1,1)=\frac{1}{\sqrt{6}} &\\
T_0(0,1,0,1)=\frac{1}{\sqrt{6}} &\\
T_0(0,1,1,0)=\frac{1}{\sqrt{6}} &\\
T_0(1,0,0,1)=\frac{1}{\sqrt{6}} &\\
T_0(1,0,1,0)=\frac{1}{\sqrt{6}} &\\
T_0(1,1,0,0)=\frac{1}{\sqrt{6}} &\\ \hline
Sz=-1  &  Sz=-2  \\ \hline
T_{-1}(0,1,1,1)=\frac{1}{2}	& T_{-2}(1,1,1,1)=1 \\
T_{-1}(1,0,1,1)=\frac{1}{2} & \\
T_{-1}(1,1,0,1)=\frac{1}{2}& \\
T_{-1}(1,1,1,0)=\frac{1}{2}& \\ \hline\hline
\caption{The $S=2$ AKLT state. $V=1/2$, $n_{occ.}= \{4\}$, $S=2$, ${\text{Pg}} = A_1$.}
\label{Table:s2AKLT}
\end{longtable}

\subsection{The spin-$1$ featureless paramagnet} 
\begin{longtable}{MMM}\hline \hline
Sz=+1 & Sz=0 & Sz=-1 \\ \hline\endfirsthead
\hline \multicolumn{2}{c}{Table \ref{Table:FPB1} (continued)} \\ \hline\endhead 
T_1(0,0,0,1)=\frac{1}{2}& T_0(0,1,0,1)=\frac{1}{\sqrt{2}} & T_{-1}(0,1,1,1)=\frac{1}{2} \\ 
T_1(0,0,1,0)=-\frac{1}{2} & T_0(1,0,1,0)=-\frac{1}{\sqrt{2}} & T_{-1}(1,0,1,1)=-\frac{1}{2}\\ 
T_1(0,1,0,0)=\frac{1}{2} & & T_{-1}(1,1,0,1)=\frac{1}{2} \\ 
T_1(1,0,0,0)=-\frac{1}{2} & & T_{-1}(1,1,1,0)=-\frac{1}{2} \\ \hline\hline
\caption{The tensor of the spin-$1$ featureless paramagnet. $V=1/2$, $n_{occ.}= \{4\}$, $S=1$, ${\text{Pg}} = B_1$.}
\label{Table:FPB1}
\end{longtable}

\subsection{The spin-$\frac{1}{2}$ Resonating Valence Bond state~: the gauge-equivalent
${\cal A}_1^{(1)}$ and ${\cal B}_1^{(1)}$ tensors} 
\begin{longtable}{NN}\hline \hline
Sz=+1/2 & Sz=-1/2   \\ \hline\endfirsthead
\hline \multicolumn{2}{c}{Table \ref{Table:RVBA1} (continued)} \\ \hline\endhead 
T_{\frac{1}{2}}(0,2,2,2)=\frac{1}{2} & T_{-\frac{1}{2}}(1,2,2,2)=\frac{1}{2} \\  
T_{\frac{1}{2}}(2,0,2,2)=\frac{1}{2} & T_{-\frac{1}{2}}(2,1,2,2)=\frac{1}{2} \\ 
T_{\frac{1}{2}}(2,2,0,2)=\frac{1}{2} & T_{-\frac{1}{2}}(2,2,1,2)=\frac{1}{2} \\ 
T_{\frac{1}{2}}(2,2,2,0)=\frac{1}{2} & T_{-\frac{1}{2}}(2,2,2,1)=\frac{1}{2} \\ 
\hline \hline
\caption{The ${\cal A}_1^{(1)}$ tensor of the NN Resonating Valence Bond (NN RVB) state. $V=1/2 \oplus 0$, $n_{occ.}= \{1,3\}$, $S=1/2$, ${\text{Pg}} = A_1$.}
\label{Table:RVBA1}
\end{longtable}

\begin{longtable}{NN}\hline \hline
Sz=+1/2 & Sz=+-1/2   \\ \hline\endfirsthead
\hline \multicolumn{2}{c}{Table \ref{Table:RVBB1} (continued)} \\ \hline\endhead
T_{\frac{1}{2}}(0,2,2,2)=\frac{1}{2} & T_{-\frac{1}{2}}(1,2,2,2)=\frac{1}{2} \\  
T_{\frac{1}{2}}(2,0,2,2)=-\frac{1}{2} & T_{-\frac{1}{2}}(2,1,2,2)=-\frac{1}{2} \\ 
T_{\frac{1}{2}}(2,2,0,2)=\frac{1}{2} & T_{-\frac{1}{2}}(2,2,1,2)=\frac{1}{2} \\ 
T_{\frac{1}{2}}(2,2,2,0)=-\frac{1}{2}  & T_{-\frac{1}{2}}(2,2,2,1)=-\frac{1}{2} \\ 
\hline \hline
\caption{The ${\cal B}_1^{(1)}$ tensor of the NN Resonating Valence Bond (NN RVB) state. $V=1/2 \oplus 0$, $n_{occ.}= \{1,3\}$, $S=1/2$, ${\text{Pg}} = B_1$. }
\label{Table:RVBB1}
\end{longtable}

\subsection{The gauge-equivalent ${\cal A}_1^{(2)}$ and ${\cal B}_1^{(2)}$ RVB tensors} 
\begin{longtable}{NN}\hline \hline
Sz=+1/2 & Sz=-1/2   \\ \hline\endfirsthead
\hline \multicolumn{2}{c}{Table \ref{Table:LRRVBA1} (continued)} \\ \hline\endhead 
T_{\frac{1}{2}}(0,0,1,2)=-\frac{1}{2 \sqrt{6}} & T_{-\frac{1}{2}}(0,1,1,2)=\frac{1}{2 \sqrt{6}} \\  
T_{\frac{1}{2}}(0,0,2,1)=-\frac{1}{2 \sqrt{6}} & T_{-\frac{1}{2}}(0,1,2,1)=-\frac{1}{\sqrt{6}}\\  
T_{\frac{1}{2}}(0,1,0,2)=\frac{1}{\sqrt{6}} & T_{-\frac{1}{2}}(0,2,1,1)=\frac{1}{2 \sqrt{6}}\\ 
T_{\frac{1}{2}}(0,1,2,0)=-\frac{1}{2 \sqrt{6}} & T_{-\frac{1}{2}}(1,0,1,2)=-\frac{1}{\sqrt{6}}\\ 
T_{\frac{1}{2}}(0,2,0,1)=\frac{1}{\sqrt{6}} & T_{-\frac{1}{2}}(1,0,2,1)=\frac{1}{2 \sqrt{6}}\\ 
T_{\frac{1}{2}}(0,2,1,0)=-\frac{1}{2 \sqrt{6}} & T_{-\frac{1}{2}}(1,1,0,2)=\frac{1}{2 \sqrt{6}}\\  
T_{\frac{1}{2}}(1,0,0,2)=-\frac{1}{2 \sqrt{6}} & T_{-\frac{1}{2}}(1,1,2,0)=\frac{1}{2 \sqrt{6}}\\  
T_{\frac{1}{2}}(1,0,2,0)=\frac{1}{\sqrt{6}} & T_{-\frac{1}{2}}(1,2,0,1)=\frac{1}{2 \sqrt{6}}\\  
T_{\frac{1}{2}}(1,2,0,0)=-\frac{1}{2 \sqrt{6}} & T_{-\frac{1}{2}}(1,2,1,0)=-\frac{1}{\sqrt{6}}\\ 
T_{\frac{1}{2}}(2,0,0,1)=-\frac{1}{2 \sqrt{6}} & T_{-\frac{1}{2}}(2,0,1,1)=\frac{1}{2 \sqrt{6}}\\ 
T_{\frac{1}{2}}(2,0,1,0)=\frac{1}{\sqrt{6}} & T_{-\frac{1}{2}}(2,1,0,1)=-\frac{1}{\sqrt{6}}\\ 
T_{\frac{1}{2}}(2,1,0,0)=-\frac{1}{2 \sqrt{6}}  & T_{-\frac{1}{2}}(2,1,1,0)=\frac{1}{2 \sqrt{6}} \\   
\hline \hline
\caption{The ${\cal A}_1^{(2)}$ RVB tensor. $V=1/2 \oplus 0$, $n_{occ.}= \{3,1\}$, $S=1/2$, ${\text{Pg}} = A_1$. }
\label{Table:LRRVBA1}
\end{longtable}

\begin{longtable}{NN}\hline \hline
Sz=+1/2 & Sz=-1/2   \\ \hline\endfirsthead
\hline \multicolumn{2}{c}{Table \ref{Table:LRRVBB1} (continued)} \\ \hline\endhead 
T_{\frac{1}{2}}(0,0,1,2)=-\frac{1}{2 \sqrt{6}} & T_{-\frac{1}{2}}(0,1,1,2)=\frac{1}{2 \sqrt{6}}\\ 
T_{\frac{1}{2}}(0,0,2,1)=\frac{1}{2 \sqrt{6}} & T_{-\frac{1}{2}}(0,1,2,1)=\frac{1}{\sqrt{6}}\\ 
T_{\frac{1}{2}}(0,1,0,2)=\frac{1}{\sqrt{6}} & T_{-\frac{1}{2}}(0,2,1,1)=\frac{1}{2 \sqrt{6}}\\ 
T_{\frac{1}{2}}(0,1,2,0)=\frac{1}{2 \sqrt{6}} & T_{-\frac{1}{2}}(1,0,1,2)=-\frac{1}{\sqrt{6}}\\ 
T_{\frac{1}{2}}(0,2,0,1)=\frac{1}{\sqrt{6}} & T_{-\frac{1}{2}}(1,0,2,1)=-\frac{1}{2 \sqrt{6}}\\ 
T_{\frac{1}{2}}(0,2,1,0)=-\frac{1}{2 \sqrt{6}} & T_{-\frac{1}{2}}(1,1,0,2)=\frac{1}{2 \sqrt{6}}\\ 
T_{\frac{1}{2}}(1,0,0,2)=-\frac{1}{2 \sqrt{6}} & T_{-\frac{1}{2}}(1,1,2,0)=-\frac{1}{2 \sqrt{6}}\\ 
T_{\frac{1}{2}}(1,0,2,0)=-\frac{1}{\sqrt{6}} & T_{-\frac{1}{2}}(1,2,0,1)=\frac{1}{2 \sqrt{6}}\\ 
T_{\frac{1}{2}}(1,2,0,0)=-\frac{1}{2 \sqrt{6}} & T_{-\frac{1}{2}}(1,2,1,0)=-\frac{1}{\sqrt{6}}\\ 
T_{\frac{1}{2}}(2,0,0,1)=\frac{1}{2 \sqrt{6}} & T_{-\frac{1}{2}}(2,0,1,1)=-\frac{1}{2 \sqrt{6}}\\ 
T_{\frac{1}{2}}(2,0,1,0)=-\frac{1}{\sqrt{6}} & T_{-\frac{1}{2}}(2,1,0,1)=\frac{1}{\sqrt{6}}\\  
T_{\frac{1}{2}}(2,1,0,0)=\frac{1}{2 \sqrt{6}} & T_{-\frac{1}{2}}(2,1,1,0)=-\frac{1}{2 \sqrt{6}} \\   
\hline \hline
\caption{The ${\cal B}_1^{(2)}$ RVB tensor. $V=1/2 \oplus 0$, $n_{occ.}= \{3,1\}$, $S=1/2$, ${\text{Pg}} = B_1$.}
\label{Table:LRRVBB1}
\end{longtable}

%\clearpage
\subsection{The spin-$\frac{1}{2}$ NN fermionic RVB state (NN fRVB)} 
\begin{longtable}{NN}\hline \hline
Sz=+1/2 & Sz=-1/2   \\ \hline\endfirsthead
\hline \multicolumn{2}{c}{Table \ref{Table:NNferm} (continued)} \\ \hline\endhead 
T_{\frac{1}{2}}(0,2,2,2)=\frac{1}{2} & T_{-\frac{1}{2}}(1,2,2,2)=\frac{1}{2}\\ 
T_{\frac{1}{2}}(2,0,2,2)=\pm\frac{i}{2} & T_{-\frac{1}{2}}(2,1,2,2)=\pm\frac{i}{2}\\ 
T_{\frac{1}{2}}(2,2,0,2)=-\frac{1}{2} &  T_{-\frac{1}{2}}(2,2,1,2)=-\frac{1}{2}\\ 
T_{\frac{1}{2}}(2,2,2,0)=\mp\frac{i}{2}  & T_{-\frac{1}{2}}(2,2,2,1)=\mp\frac{i}{2}  \\   
\hline \hline
\caption{The spin-$\frac{1}{2}$ NN fermionic RVB (NN fRVB) tensor. $V=1/2 \oplus 0$, $n_{occ.}= \{1,3\}$, $S=1/2$, ${\text{Pg}} = E$.}
\label{Table:NNferm}
\end{longtable}

\subsection{The gauge-equivalent ${\cal A}_2$ and ${\cal B}_2$ RVB tensors}
\begin{longtable}{NN}\hline \hline
Sz=+1/2 & Sz=-1/2   \\ \hline\endfirsthead
\hline \multicolumn{2}{c}{Table \ref{Table:LRRVBA2} (continued)} \\ \hline\endhead 
T_{\frac{1}{2}}(0,0,1,2)=\frac{1}{2 \sqrt{2}}&    T_{-\frac{1}{2}}(0,1,1,2)=\frac{1}{2 \sqrt{2}}\\ 
T_{\frac{1}{2}}(0,0,2,1)=-\frac{1}{2 \sqrt{2}}&   T_{-\frac{1}{2}}(0,2,1,1)=-\frac{1}{2 \sqrt{2}}\\
T_{\frac{1}{2}}(0,1,2,0)=\frac{1}{2 \sqrt{2}}&    T_{-\frac{1}{2}}(1,0,2,1)=-\frac{1}{2 \sqrt{2}}\\
T_{\frac{1}{2}}(0,2,1,0)=-\frac{1}{2 \sqrt{2}}&   T_{-\frac{1}{2}}(1,1,0,2)=-\frac{1}{2 \sqrt{2}}\\
T_{\frac{1}{2}}(1,0,0,2)=-\frac{1}{2 \sqrt{2}}&   T_{-\frac{1}{2}}(1,1,2,0)=\frac{1}{2 \sqrt{2}}\\ 
T_{\frac{1}{2}}(1,2,0,0)=\frac{1}{2 \sqrt{2}}&    T_{-\frac{1}{2}}(1,2,0,1)=\frac{1}{2 \sqrt{2}}\\ 
T_{\frac{1}{2}}(2,0,0,1)=\frac{1}{2 \sqrt{2}}&    T_{-\frac{1}{2}}(2,0,1,1)=\frac{1}{2 \sqrt{2}}\\ 
T_{\frac{1}{2}}(2,1,0,0)=-\frac{1}{2 \sqrt{2}}&   T_{-\frac{1}{2}}(2,1,1,0)=-\frac{1}{2 \sqrt{2}}\\
\hline \hline
\caption{The ${\cal A}_2$ RVB tensor. $V=1/2 \oplus 0$, $n_{occ.}= \{3,1\}$, $S=1/2$, ${\text{Pg}} = A_2$. }
\label{Table:LRRVBA2}
\end{longtable}

\begin{longtable}{NN}\hline \hline
Sz=+1/2 & Sz=-1/2   \\ \hline\endfirsthead
\hline \multicolumn{2}{c}{Table \ref{Table:LRRVBB2} (continued)} \\ \hline\endhead 
T_{\frac{1}{2}}(0,0,1,2)=\frac{1}{2 \sqrt{2}}&       T_{-\frac{1}{2}}(0,1,1,2)=\frac{1}{2 \sqrt{2}}\\   
T_{\frac{1}{2}}(0,0,2,1)=\frac{1}{2 \sqrt{2}}&       T_{-\frac{1}{2}}(0,2,1,1)=-\frac{1}{2 \sqrt{2}}\\  
T_{\frac{1}{2}}(0,1,2,0)=-\frac{1}{2 \sqrt{2}}&      T_{-\frac{1}{2}}(1,0,2,1)=\frac{1}{2 \sqrt{2}}\\   
T_{\frac{1}{2}}(0,2,1,0)=-\frac{1}{2 \sqrt{2}}&      T_{-\frac{1}{2}}(1,1,0,2)=-\frac{1}{2 \sqrt{2}}\\  
T_{\frac{1}{2}}(1,0,0,2)=-\frac{1}{2 \sqrt{2}}&      T_{-\frac{1}{2}}(1,1,2,0)=-\frac{1}{2 \sqrt{2}}\\  
T_{\frac{1}{2}}(1,2,0,0)=\frac{1}{2 \sqrt{2}}&       T_{-\frac{1}{2}}(1,2,0,1)=\frac{1}{2 \sqrt{2}}\\   
T_{\frac{1}{2}}(2,0,0,1)=-\frac{1}{2 \sqrt{2}}&      T_{-\frac{1}{2}}(2,0,1,1)=-\frac{1}{2 \sqrt{2}}\\  
T_{\frac{1}{2}}(2,1,0,0)=\frac{1}{2 \sqrt{2}}&       T_{-\frac{1}{2}}(2,1,1,0)=\frac{1}{2 \sqrt{2}}\\   
\hline \hline
\caption{The ${\cal A}_2$ RVB tensor. $V=1/2 \oplus 0$, $n_{occ.}= \{3,1\}$, $S=1/2$, ${\text{Pg}} = B_2$. }
\label{Table:LRRVBB2}
\end{longtable}

\subsection{The spin-$1$ Resonating Loop (RAL) State} 
\begin{longtable}{MMM}\hline \hline
Sz=+1 & Sz=0 & Sz=-1 \\ \hline\endfirsthead
\hline \multicolumn{2}{c}{Table \ref{Table:RAL} (continued)} \\ \hline\endhead 
T_1^{(a)}(0,0,2,2)=\frac{1}{2} & T_0^{(a)}(0,1,2,2)=\frac{1}{2 \sqrt{2}} & T_{-1}^{(a)}(1,1,2,2)=\frac{1}{2}\\ 
T_1^{(a)}(0,2,2,0)=\frac{1}{2} & T_0^{(a)}(0,2,2,1)=\frac{1}{2 \sqrt{2}} & T_{-1}^{(a)}(1,2,2,1)=\frac{1}{2}\\ 
T_1^{(a)}(2,0,0,2)=\frac{1}{2} & T_0^{(a)}(1,0,2,2)=\frac{1}{2 \sqrt{2}} & T_{-1}^{(a)}(2,1,1,2)=\frac{1}{2}\\ 
T_1^{(a)}(2,2,0,0)=\frac{1}{2} & T_0^{(a)}(1,2,2,0)=\frac{1}{2 \sqrt{2}} & T_{-1}^{(a)}(2,2,1,1)=\frac{1}{2}\\  
 & T_0^{(a)}(2,0,1,2)=\frac{1}{2 \sqrt{2}}& \\ 
 & T_0^{(a)}(2,1,0,2)=\frac{1}{2 \sqrt{2}}& \\ 
 & T_0^{(a)}(2,2,0,1)=\frac{1}{2 \sqrt{2}}& \\ 
 & T_0^{(a)}(2,2,1,0)=\frac{1}{2 \sqrt{2}}& \\
 \hline 
Sz=+1 & Sz=0 & Sz=-1 \\ \hline
T_1^{(b)}(0,2,0,2)=\frac{1}{\sqrt{2}} & T_0^{(b)}(0,2,1,2)=\frac{1}{2}& T_{-1}^{(b)}(1,2,1,2)=\frac{1}{\sqrt{2}}\\ 
T_1^{(b)}(2,0,2,0)=\frac{1}{\sqrt{2}} & T_0^{(b)}(1,2,0,2)=\frac{1}{2}& T_{-1}^{(b)}(2,1,2,1)=\frac{1}{\sqrt{2}}\\ 
 & T_0^{(b)}(2,0,2,1)=\frac{1}{2} & \\ 
 & T_0^{(b)}(2,1,2,0)=\frac{1}{2} & \\  \hline\hline
\caption{The tensors of the spin-$1$ Resonating Loop (RAL) State. $V=1/2 \oplus 0$, $n_{occ.}= \{2,2\}$, $S=1$, ${\text{Pg}} = A_1$.}
\label{Table:RAL}
\end{longtable}

\subsection{The generalized spin-$S$ NN RVB state} 
\begin{longtable}{MMM}\hline \hline
Sz=+1 & Sz=0 & Sz=-1 \\ \hline\endfirsthead
\hline \multicolumn{2}{c}{Table \ref{Table:GNNRVB1A1} (continued)} \\ \hline\endhead 
T_1(0,3,3,3)=\frac{1}{2} & T_0(1,3,3,3)=\frac{1}{2} & T_{-1}(2,3,3,3)=\frac{1}{2} \\ 
T_1(3,0,3,3)=\frac{1}{2} & T_0(3,1,3,3)=\frac{1}{2} & T_{-1}(3,2,3,3)=\frac{1}{2} \\ 
T_1(3,3,0,3)=\frac{1}{2} & T_0(3,3,1,3)=\frac{1}{2} & T_{-1}(3,3,2,3)=\frac{1}{2} \\ 
T_1(3,3,3,0)=\frac{1}{2} & T_0(3,3,3,1)=\frac{1}{2} & T_{-1}(3,3,3,2)=\frac{1}{2} \\ \hline\hline
\caption{The tensor of the generalized spin-$1$ NN RVB state. $V=1 \oplus 0$, $n_{occ.}= \{1,3\}$, $S=1$, ${\text{Pg}} = A_1$.}
\label{Table:GNNRVB1A1}
\end{longtable}

\begin{longtable}{MMM}\hline \hline
Sz=+1 & Sz=0 & Sz=-1 \\ \hline\endfirsthead
\hline \multicolumn{2}{c}{Table \ref{Table:GNNRVB1B1} (continued)} \\ \hline\endhead 
T_1(0,3,3,3)=\frac{1}{2} & T_0(1,3,3,3)=\frac{1}{2} & T_{-1}(2,3,3,3)=\frac{1}{2}\\ 
T_1(3,0,3,3)=-\frac{1}{2} & T_0(3,1,3,3)=-\frac{1}{2} & T_{-1}(3,2,3,3)=-\frac{1}{2}\\ 
T_1(3,3,0,3)=\frac{1}{2} & T_0(3,3,1,3)=\frac{1}{2} & T_{-1}(3,3,2,3)=\frac{1}{2}\\ 
T_1(3,3,3,0)=-\frac{1}{2} & T_0(3,3,3,1)=-\frac{1}{2} & T_{-1}(3,3,3,2)=-\frac{1}{2}  \\ \hline\hline
\caption{The tensor of the generalized spin-$1$ NN RVB state. $V=1 \oplus 0$, $n_{occ.}= \{1,3\}$, $S=1$, ${\text{Pg}} = B_1$.}
\label{Table:GNNRVB1B1}
\end{longtable}

\begin{longtable}{NN}\hline \hline
Sz=+3/2 & Sz=+1/2   \\ \hline\endfirsthead
\hline \multicolumn{2}{c}{Table \ref{Table:GNNRVB32A1} (continued)} \\ \hline\endhead 
T_{\frac{3}{2}}(0,4,4,4)=\frac{1}{2} & T_{\frac{1}{2}}(1,4,4,4)=\frac{1}{2} \\ 
T_{\frac{3}{2}}(4,0,4,4)=\frac{1}{2} & T_{\frac{1}{2}}(4,1,4,4)=\frac{1}{2} \\ 
T_{\frac{3}{2}}(4,4,0,4)=\frac{1}{2} & T_{\frac{1}{2}}(4,4,1,4)=\frac{1}{2} \\ 
T_{\frac{3}{2}}(4,4,4,0)=\frac{1}{2} & T_{\frac{1}{2}}(4,4,4,1)=\frac{1}{2} \\ \hline
Sz=-1/2 & Sz=-3/2 \\ \hline
T_{-\frac{1}{2}}(2,4,4,4)=\frac{1}{2} & T_{-\frac{3}{2}}(3,4,4,4)=\frac{1}{2}\\ 
T_{-\frac{1}{2}}(4,2,4,4)=\frac{1}{2} & T_{-\frac{3}{2}}(4,3,4,4)=\frac{1}{2}\\ 
T_{-\frac{1}{2}}(4,4,2,4)=\frac{1}{2} & T_{-\frac{3}{2}}(4,4,3,4)=\frac{1}{2}\\ 
T_{-\frac{1}{2}}(4,4,4,2)=\frac{1}{2} & T_{-\frac{3}{2}}(4,4,4,3)=\frac{1}{2} \\ \hline\hline
\caption{The tensor of the generalized spin-$\frac{3}{2}$ NN RVB state. $V=3/2 \oplus 0$, $n_{occ.}= \{1,3\}$, $S=3/2$, ${\text{Pg}} = A_1$.}
\label{Table:GNNRVB32A1}
\end{longtable}

\begin{longtable}{NN}\hline \hline
Sz=+3/2 & Sz=+1/2   \\ \hline\endfirsthead
\hline \multicolumn{2}{c}{Table \ref{Table:GNNRVB32B1} (continued)} \\ \hline\endhead 
T_{\frac{3}{2}}(0,4,4,4)=\frac{1}{2} & T_{\frac{1}{2}}(1,4,4,4)=\frac{1}{2} \\ 
T_{\frac{3}{2}}(4,0,4,4)=-\frac{1}{2} & T_{\frac{1}{2}}(4,1,4,4)=-\frac{1}{2} \\ 
T_{\frac{3}{2}}(4,4,0,4)=\frac{1}{2} & T_{\frac{1}{2}}(4,4,1,4)=\frac{1}{2} \\ 
T_{\frac{3}{2}}(4,4,4,0)=-\frac{1}{2}  & T_{\frac{1}{2}}(4,4,4,1)=-\frac{1}{2} \\ \hline
Sz=-1/2 & Sz=-3/2 \\ \hline
T_{-\frac{1}{2}}(2,4,4,4)=\frac{1}{2} & T_{-\frac{3}{2}}(3,4,4,4)=\frac{1}{2}\\ 
T_{-\frac{1}{2}}(4,2,4,4)=-\frac{1}{2} & T_{-\frac{3}{2}}(4,3,4,4)=-\frac{1}{2}\\ 
T_{-\frac{1}{2}}(4,4,2,4)=\frac{1}{2} & T_{-\frac{3}{2}}(4,4,3,4)=\frac{1}{2}\\ 
T_{-\frac{1}{2}}(4,4,4,2)=-\frac{1}{2} & T_{-\frac{3}{2}}(4,4,4,3)=-\frac{1}{2} \\ \hline\hline
\caption{The tensor of the generalized spin-$\frac{3}{2}$ NN RVB state. $V=3/2 \oplus 0$, $n_{occ.}= \{1,3\}$, $S=3/2$, ${\text{Pg}} = B_1$.}
\label{Table:GNNRVB32B1}
\end{longtable}

%\clearpage
\section{$S$ and $G$ tensors obtained from the decomposition of the double-layer CSL}
Here one uses the same set of indices for all tensors, according to Fig.~\ref{FIG:single_double}(b).
\label{app3}

% S1
\begin{longtable}{LLL}\hline \hline
Sz=+1 & Sz=0  & Sz=-1  \\ \hline\endfirsthead 
\hline Sz=+1 \text{ (cont.)} & Sz=0 \text{ (cont.)}& Sz=-1 \text{ (cont.)} \\ \hline\endhead 
T_1(2,2,6,7)=\frac{1}{4 \sqrt{3}}& T_0(2,2,7,7)=\frac{1}{2 \sqrt{6}}&T_{-1}(2,5,7,7)=\frac{1}{4 \sqrt{3}}\\
T_1(2,2,7,6)=\frac{1}{4 \sqrt{3}}& T_0(2,7,2,7)=-\frac{1}{\sqrt{6}}& T_{-1}(2,7,5,7)=-\frac{1}{2 \sqrt{3}}\\
T_1(2,5,6,6)=-\frac{1}{4 \sqrt{3}}&T_0(2,7,7,2)=\frac{1}{2 \sqrt{6}}&T_{-1}(2,7,7,5)=\frac{1}{4 \sqrt{3}}\\ 
T_1(2,6,2,7)=-\frac{1}{2 \sqrt{3}}&T_0(5,5,6,6)=-\frac{1}{2 \sqrt{6}}&T_{-1}(5,2,7,7)=\frac{1}{4 \sqrt{3}}\\ 
T_1(2,6,5,6)=\frac{1}{2 \sqrt{3}}& T_0(5,6,5,6)=\frac{1}{\sqrt{6}}&  T_{-1}(5,5,6,7)=-\frac{1}{4 \sqrt{3}}\\
T_1(2,6,6,5)=-\frac{1}{4 \sqrt{3}}&T_0(5,6,6,5)=-\frac{1}{2 \sqrt{6}}&T_{-1}(5,5,7,6)=-\frac{1}{4 \sqrt{3}}\\ 
T_1(2,6,7,2)=\frac{1}{4 \sqrt{3}}& T_0(6,5,5,6)=-\frac{1}{2 \sqrt{6}}&T_{-1}(5,6,5,7)=\frac{1}{2 \sqrt{3}}\\
T_1(2,7,2,6)=-\frac{1}{2 \sqrt{3}}&T_0(6,5,6,5)=\frac{1}{\sqrt{6}}&  T_{-1}(5,6,7,5)=-\frac{1}{4 \sqrt{3}}\\
\end{longtable}

% G1
\begin{longtable}{LLL}\hline \hline
Sz=+1 & Sz=0  & Sz=-1  \\ \hline\endfirsthead 
\hline Sz=+1 \text{ (cont.)} & Sz=0 \text{ (cont.)}& Sz=-1 \text{ (cont.)} \\ \hline\endhead 
T_1(2,2,6,7)=-\frac{1}{4}&    T_0(2,5,7,6)=\frac{1}{2 \sqrt{2}}&    T_{-1}(2,5,7,7)=\frac{1}{4}\\ 
T_1(2,2,7,6)=\frac{1}{4}&     T_0(2,6,7,5)=-\frac{1}{2 \sqrt{2}}&   T_{-1}(2,7,7,5)=-\frac{1}{4}\\
T_1(2,5,6,6)=\frac{1}{4}&     T_0(5,2,6,7)=-\frac{1}{2 \sqrt{2}}&   T_{-1}(5,2,7,7)=-\frac{1}{4}\\
T_1(2,6,6,5)=-\frac{1}{4}&    T_0(5,7,6,2)=\frac{1}{2 \sqrt{2}}&    T_{-1}(5,5,6,7)=-\frac{1}{4}\\
T_1(2,6,7,2)=-\frac{1}{4}&    T_0(6,2,5,7)=\frac{1}{2 \sqrt{2}}&    T_{-1}(5,5,7,6)=\frac{1}{4}\\ 
T_1(2,7,6,2)=\frac{1}{4}&     T_0(6,7,5,2)=-\frac{1}{2 \sqrt{2}}&   T_{-1}(5,6,7,5)=-\frac{1}{4}\\
T_1(5,2,6,6)=-\frac{1}{4}&    T_0(7,5,2,6)=-\frac{1}{2 \sqrt{2}}&   T_{-1}(5,7,6,5)=\frac{1}{4}\\ 
T_1(5,6,6,2)=\frac{1}{4}&     T_0(7,6,2,5)=\frac{1}{2 \sqrt{2}}&      T_{-1}(5,7,7,2)=\frac{1}{4}\\ 
T_1(6,2,2,7)=\frac{1}{4}&&                                             T_{-1}(6,5,5,7)=\frac{1}{4}\\ 
T_1(6,2,5,6)=\frac{1}{4}&&                                             T_{-1}(6,7,5,5)=-\frac{1}{4}\\
T_1(6,5,2,6)=-\frac{1}{4}&&                                            T_{-1}(7,2,5,7)=\frac{1}{4}\\ 
T_1(6,6,2,5)=\frac{1}{4}&&                                             T_{-1}(7,5,2,7)=-\frac{1}{4}\\
T_1(6,6,5,2)=-\frac{1}{4}&&                                            T_{-1}(7,5,5,6)=-\frac{1}{4}\\
T_1(6,7,2,2)=-\frac{1}{4}&&                                            T_{-1}(7,6,5,5)=\frac{1}{4}\\ 
T_1(7,2,2,6)=-\frac{1}{4}&&                                           T_{-1}(7,7,2,5)=\frac{1}{4}\\ 
T_1(7,6,2,2)=\frac{1}{4}&&                                               T_{-1}(7,7,5,2)=-\frac{1}{4}  \\
\hline\hline
\caption{$G_1$.}
\label{Table:G1}
\end{longtable}

% S2
\begin{longtable}{LLL}\hline \hline
Sz=+1 & Sz=0  & Sz=-1  \\ \hline\endfirsthead 
\hline Sz=+1 \text{ (cont.)} & Sz=0 \text{ (cont.)}& Sz=-1 \text{ (cont.)} \\ \hline\endhead 
T_1(2,2,8,8)=\frac{1}{2 \sqrt{3}}&     T_0(2,5,8,8)=\frac{1}{2 \sqrt{6}}&     T_{-1}(5,5,8,8)=\frac{1}{2 \sqrt{3}}\\ 
T_1(2,8,2,8)=\frac{1}{2 \sqrt{3}}&     T_0(2,8,5,8)=\frac{1}{2 \sqrt{6}}&     T_{-1}(5,8,5,8)=\frac{1}{2 \sqrt{3}}\\ 
T_1(2,8,8,2)=\frac{1}{2 \sqrt{3}}&     T_0(2,8,8,5)=\frac{1}{2 \sqrt{6}}&     T_{-1}(5,8,8,5)=\frac{1}{2 \sqrt{3}}\\ 
T_1(6,6,8,8)=-\frac{1}{2 \sqrt{3}}&    T_0(5,2,8,8)=\frac{1}{2 \sqrt{6}}&     T_{-1}(7,7,8,8)=-\frac{1}{2 \sqrt{3}}\\
T_1(6,8,6,8)=-\frac{1}{2 \sqrt{3}}&    T_0(5,8,2,8)=\frac{1}{2 \sqrt{6}}&     T_{-1}(7,8,7,8)=-\frac{1}{2 \sqrt{3}}\\
T_1(6,8,8,6)=-\frac{1}{2 \sqrt{3}}&    T_0(5,8,8,2)=\frac{1}{2 \sqrt{6}}&     T_{-1}(7,8,8,7)=-\frac{1}{2 \sqrt{3}}\\
T_1(8,2,2,8)=\frac{1}{2 \sqrt{3}}&     T_0(6,7,8,8)=-\frac{1}{2 \sqrt{6}}&    T_{-1}(8,5,5,8)=\frac{1}{2 \sqrt{3}}\\ 
T_1(8,2,8,2)=\frac{1}{2 \sqrt{3}}&     T_0(6,8,7,8)=-\frac{1}{2 \sqrt{6}}&    T_{-1}(8,5,8,5)=\frac{1}{2 \sqrt{3}}\\
\end{longtable}

% S3
\begin{longtable}{LLL}\hline \hline
Sz=+1 & Sz=0  & Sz=-1  \\ \hline\endfirsthead 
\hline Sz=+1 \text{ (cont.)} & Sz=0 \text{ (cont.)}& Sz=-1 \text{ (cont.)} \\ \hline\endhead 
T_1(0,8,8,8)=\frac{1}{2}&      T_0(1,8,8,8)=\frac{1}{2}&      T_{-1}(4,8,8,8)=\frac{1}{2}\\   
T_1(8,0,8,8)=\frac{1}{2}&       T_0(8,1,8,8)=\frac{1}{2}&      T_{-1}(8,4,8,8)=\frac{1}{2}\\   
T_1(8,8,0,8)=\frac{1}{2}&       T_0(8,8,1,8)=\frac{1}{2}&      T_{-1}(8,8,4,8)=\frac{1}{2}\\   
T_1(8,8,8,0)=\frac{1}{2}&         T_0(8,8,8,1)=\frac{1}{2} &       T_{-1}(8,8,8,4)=\frac{1}{2} \\    
\hline\hline
\caption{$S_3$.}
\label{Table:S3}
\end{longtable}

% S4a
\begin{longtable}{LLL}\hline \hline
Sz=+1 & Sz=0  & Sz=-1  \\ \hline\endfirsthead 
\hline Sz=+1 \text{ (cont.)} & Sz=0 \text{ (cont.)}& Sz=-1 \text{ (cont.)} \\ \hline\endhead 
T_1(0,0,4,8)=\frac{1}{4 \sqrt{2}}&       T_0(0,1,8,4)=\frac{1}{4 \sqrt{2}}&    T_{-1}(0,4,4,8)=\frac{1}{4 \sqrt{2}}\\ 
T_1(0,0,8,4)=\frac{1}{4 \sqrt{2}}&       T_0(0,4,1,8)=\frac{1}{4 \sqrt{2}}&    T_{-1}(0,4,8,4)=\frac{1}{2 \sqrt{2}}\\ 
T_1(0,1,1,8)=-\frac{1}{4 \sqrt{2}}&      T_0(0,4,8,1)=\frac{1}{4 \sqrt{2}}&    T_{-1}(0,8,4,4)=\frac{1}{4 \sqrt{2}}\\ 
T_1(0,4,0,8)=\frac{1}{2 \sqrt{2}}&       T_0(0,8,1,4)=\frac{1}{4 \sqrt{2}}&    T_{-1}(1,1,4,8)=-\frac{1}{4 \sqrt{2}}\\ 
T_1(0,4,8,0)=\frac{1}{4 \sqrt{2}}&       T_0(1,0,4,8)=\frac{1}{4 \sqrt{2}}&    T_{-1}(1,1,8,4)=-\frac{1}{4 \sqrt{2}}\\ 
T_1(0,8,0,4)=\frac{1}{2 \sqrt{2}}&       T_0(1,1,1,8)=-\frac{1}{2 \sqrt{2}}&   T_{-1}(1,4,8,1)=-\frac{1}{4 \sqrt{2}}\\ 
T_1(0,8,1,1)=-\frac{1}{4 \sqrt{2}}&      T_0(1,1,8,1)=-\frac{1}{2 \sqrt{2}}&   T_{-1}(1,8,4,1)=-\frac{1}{4 \sqrt{2}}\\ 
T_1(0,8,4,0)=\frac{1}{4 \sqrt{2}}&       T_0(1,4,0,8)=\frac{1}{4 \sqrt{2}}&    T_{-1}(4,0,4,8)=\frac{1}{2 \sqrt{2}}\\
\end{longtable} 

%S4b
\begin{longtable}{LLL}\hline \hline
Sz=+1 & Sz=0  & Sz=-1  \\ \hline\endfirsthead 
\hline Sz=+1 \text{ (cont.)} & Sz=0 \text{ (cont.)}& Sz=-1 \text{ (cont.)} \\ \hline\endhead 
T_1(0,0,4,8)=-\frac{3}{4 \sqrt{10}}&      T_0(0,1,4,8)=-\frac{1}{\sqrt{10}}&  T_{-1}(0,4,4,8)=-\frac{3}{4 \sqrt{10}}\\   
T_1(0,0,8,4)=-\frac{3}{4 \sqrt{10}}&  T_0(0,1,8,4)=\frac{1}{4 \sqrt{10}}&     T_{-1}(0,4,8,4)=\frac{1}{2 \sqrt{10}}\\   
T_1(0,1,1,8)=-\frac{1}{4 \sqrt{10}}&  T_0(0,4,1,8)=\frac{1}{4 \sqrt{10}}&     T_{-1}(0,8,4,4)=-\frac{3}{4 \sqrt{10}}\\  
T_1(0,1,8,1)=\frac{1}{\sqrt{10}}&     T_0(0,4,8,1)=\frac{1}{4 \sqrt{10}}&     T_{-1}(1,1,4,8)=-\frac{1}{4 \sqrt{10}}\\  
T_1(0,4,0,8)=\frac{1}{2 \sqrt{10}}&   T_0(0,8,1,4)=\frac{1}{4 \sqrt{10}}&     T_{-1}(1,1,8,4)=-\frac{1}{4 \sqrt{10}}\\  
T_1(0,4,8,0)=-\frac{3}{4 \sqrt{10}}&  T_0(0,8,4,1)=-\frac{1}{\sqrt{10}}&  T_{-1}(1,4,1,8)=\frac{1}{\sqrt{10}}\\     
T_1(0,8,0,4)=\frac{1}{2 \sqrt{10}}&   T_0(1,0,4,8)=\frac{1}{4 \sqrt{10}}&    T_{-1}(1,4,8,1)=-\frac{1}{4 \sqrt{10}}\\  
T_1(0,8,1,1)=-\frac{1}{4 \sqrt{10}}&  T_0(1,0,8,4)=-\frac{1}{\sqrt{10}}&  T_{-1}(1,8,1,4)=\frac{1}{\sqrt{10}}\\
\end{longtable}  

%G4
\begin{longtable}{LLL}\hline \hline
Sz=+1 & Sz=0  & Sz=-1  \\ \hline\endfirsthead 
\hline Sz=+1 \text{ (cont.)} & Sz=0 \text{ (cont.)}& Sz=-1 \text{ (cont.)} \\ \hline\endhead 
T_1(0,0,4,8)=\frac{1}{4}&  T_0(0,1,8,4)=-\frac{1}{4}&    T_{-1}(0,4,4,8)=-\frac{1}{4}\\ 
T_1(0,0,8,4)=-\frac{1}{4}&       T_0(0,4,1,8)=-\frac{1}{4}&    T_{-1}(0,8,4,4)=\frac{1}{4}\\  
T_1(0,1,1,8)=-\frac{1}{4}&       T_0(0,4,8,1)=\frac{1}{4}&           T_{-1}(1,1,4,8)=\frac{1}{4}\\  
T_1(0,4,8,0)=\frac{1}{4}&        T_0(0,8,1,4)=\frac{1}{4}&           T_{-1}(1,1,8,4)=-\frac{1}{4}\\ 
T_1(0,8,1,1)=\frac{1}{4}&        T_0(1,0,4,8)=\frac{1}{4}&           T_{-1}(1,4,8,1)=\frac{1}{4}\\  
T_1(0,8,4,0)=-\frac{1}{4}&       T_0(1,4,0,8)=\frac{1}{4}&           T_{-1}(1,8,4,1)=-\frac{1}{4}\\ 
T_1(1,0,8,1)=\frac{1}{4}&        T_0(1,8,0,4)=-\frac{1}{4}&          T_{-1}(4,0,8,4)=\frac{1}{4}\\  
T_1(1,1,0,8)=\frac{1}{4}&        T_0(1,8,4,0)=-\frac{1}{4}&          T_{-1}(4,1,1,8)=-\frac{1}{4}\\ 
T_1(1,1,8,0)=-\frac{1}{4}&       T_0(4,0,1,8)=-\frac{1}{4}&          T_{-1}(4,4,0,8)=\frac{1}{4}\\  
T_1(1,8,0,1)=-\frac{1}{4}&       T_0(4,0,8,1)=\frac{1}{4}&           T_{-1}(4,4,8,0)=-\frac{1}{4}\\ 
T_1(4,0,0,8)=-\frac{1}{4}&       T_0(4,1,8,0)=-\frac{1}{4}&          T_{-1}(4,8,0,4)=-\frac{1}{4}\\ 
T_1(4,8,0,0)=\frac{1}{4}&        T_0(4,8,1,0)=\frac{1}{4}&           T_{-1}(4,8,1,1)=\frac{1}{4}\\  
T_1(8,0,0,4)=\frac{1}{4}&        T_0(8,0,4,1)=-\frac{1}{4}&          T_{-1}(8,0,4,4)=-\frac{1}{4}\\ 
T_1(8,0,1,1)=-\frac{1}{4}&       T_0(8,1,0,4)=\frac{1}{4}&           T_{-1}(8,1,1,4)=\frac{1}{4}\\  
T_1(8,1,1,0)=\frac{1}{4}&        T_0(8,1,4,0)=\frac{1}{4}&           T_{-1}(8,4,1,1)=-\frac{1}{4}\\ 
T_1(8,4,0,0)=-\frac{1}{4}&       T_0(8,4,0,1)=-\frac{1}{4}&          T_{-1}(8,4,4,0)=\frac{1}{4} \\   
\hline\hline
\caption{$G_4$.}
\label{Table:G4}
\end{longtable} 

%S5a
\begin{longtable}{LLL}\hline \hline
Sz=+1 & Sz=0  & Sz=-1  \\ \hline\endfirsthead
\hline Sz=+1 & Sz=0& Sz=-1  \\ \hline\endhead 
T_1(0,3,3,8)=\frac{1}{2 \sqrt{2}}&        T_0(1,3,3,8)=\frac{1}{2 \sqrt{2}}&             T_{-1}(3,3,4,8)=\frac{1}{2 \sqrt{2}}\\
T_1(0,8,3,3)=\frac{1}{2 \sqrt{2}}&        T_0(1,8,3,3)=\frac{1}{2 \sqrt{2}}&             T_{-1}(3,3,8,4)=\frac{1}{2 \sqrt{2}}\\
T_1(3,0,8,3)=\frac{1}{2 \sqrt{2}}&        T_0(3,1,8,3)=\frac{1}{2 \sqrt{2}}&             T_{-1}(3,4,8,3)=\frac{1}{2 \sqrt{2}}\\
T_1(3,3,0,8)=\frac{1}{2 \sqrt{2}}&        T_0(3,3,1,8)=\frac{1}{2 \sqrt{2}}&             T_{-1}(3,8,4,3)=\frac{1}{2 \sqrt{2}}\\
T_1(3,3,8,0)=\frac{1}{2 \sqrt{2}}&        T_0(3,3,8,1)=\frac{1}{2 \sqrt{2}}&             T_{-1}(4,3,3,8)=\frac{1}{2 \sqrt{2}}\\
T_1(3,8,0,3)=\frac{1}{2 \sqrt{2}}&        T_0(3,8,1,3)=\frac{1}{2 \sqrt{2}}&             T_{-1}(4,8,3,3)=\frac{1}{2 \sqrt{2}}\\
T_1(8,0,3,3)=\frac{1}{2 \sqrt{2}}&        T_0(8,1,3,3)=\frac{1}{2 \sqrt{2}}&             T_{-1}(8,3,3,4)=\frac{1}{2 \sqrt{2}}\\
T_1(8,3,3,0)=\frac{1}{2 \sqrt{2}}&        T_0(8,3,3,1)=\frac{1}{2 \sqrt{2}}&             T_{-1}(8,4,3,3)=\frac{1}{2 \sqrt{2}}\\  
\hline\hline
\caption{$S_5^{(a)}$.}
\label{Table:S5a}
\end{longtable} 

%S5b
\begin{longtable}{LLL}\hline \hline
Sz=+1 & Sz=0  & Sz=-1  \\ \hline\endfirsthead 
\hline Sz=+1 \text{ (cont.)} & Sz=0 \text{ (cont.)}& Sz=-1 \text{ (cont.)} \\ \hline\endhead 
T_1(0,3,8,3)=\frac{1}{2}&      T_0(1,3,8,3)=\frac{1}{2}&     T_{-1}(3,4,3,8)=\frac{1}{2}\\ 
T_1(3,0,3,8)=\frac{1}{2}&      T_0(3,1,3,8)=\frac{1}{2}&     T_{-1}(3,8,3,4)=\frac{1}{2}\\ 
T_1(3,8,3,0)=\frac{1}{2}&      T_0(3,8,3,1)=\frac{1}{2}&     T_{-1}(4,3,8,3)=\frac{1}{2}\\ 
T_1(8,3,0,3)=\frac{1}{2}&      T_0(8,3,1,3)=\frac{1}{2}&     T_{-1}(8,3,4,3)=\frac{1}{2}\\   
\hline\hline
\caption{$S_5^{(b)}$.}
\label{Table:S5b}
\end{longtable} 

% G5
\begin{longtable}{LLL}\hline \hline
Sz=+1 & Sz=0  & Sz=-1  \\ \hline\endfirsthead 
\hline Sz=+1 \text{ (cont.)} & Sz=0 \text{ (cont.)}& Sz=-1 \text{ (cont.)} \\ \hline\endhead 
T_1(0,3,3,8)=\frac{1}{2 \sqrt{2}}&      T_0(1,3,3,8)=\frac{1}{2 \sqrt{2}}&       T_{-1}(3,3,4,8)=-\frac{1}{2 \sqrt{2}}\\  
T_1(0,8,3,3)=-\frac{1}{2 \sqrt{2}}&     T_0(1,8,3,3)=-\frac{1}{2 \sqrt{2}}&      T_{-1}(3,3,8,4)=\frac{1}{2 \sqrt{2}}\\   
T_1(3,0,8,3)=-\frac{1}{2 \sqrt{2}}&     T_0(3,1,8,3)=-\frac{1}{2 \sqrt{2}}&      T_{-1}(3,4,8,3)=-\frac{1}{2 \sqrt{2}}\\  
T_1(3,3,0,8)=-\frac{1}{2 \sqrt{2}}&     T_0(3,3,1,8)=-\frac{1}{2 \sqrt{2}}&      T_{-1}(3,8,4,3)=\frac{1}{2 \sqrt{2}}\\   
T_1(3,3,8,0)=\frac{1}{2 \sqrt{2}}&      T_0(3,3,8,1)=\frac{1}{2 \sqrt{2}}&       T_{-1}(4,3,3,8)=\frac{1}{2 \sqrt{2}}\\   
T_1(3,8,0,3)=\frac{1}{2 \sqrt{2}}&      T_0(3,8,1,3)=\frac{1}{2 \sqrt{2}}&       T_{-1}(4,8,3,3)=-\frac{1}{2 \sqrt{2}}\\  
T_1(8,0,3,3)=\frac{1}{2 \sqrt{2}}&      T_0(8,1,3,3)=\frac{1}{2 \sqrt{2}}&       T_{-1}(8,3,3,4)=-\frac{1}{2 \sqrt{2}}\\  
T_1(8,3,3,0)=-\frac{1}{2 \sqrt{2}}&     T_0(8,3,3,1)=-\frac{1}{2 \sqrt{2}}&      T_{-1}(8,4,3,3)=\frac{1}{2 \sqrt{2}}\\     
\hline\hline
\caption{$G_5$.}
\label{Table:G5}
\end{longtable} 

%S6a
\begin{longtable}{LLL}\hline \hline
Sz=+1 & Sz=0  & Sz=-1  \\ \hline\endfirsthead 
\hline Sz=+1 \text{ (cont.)} & Sz=0 \text{ (cont.)}& Sz=-1 \text{ (cont.)} \\ \hline\endhead 
T_1(2,2,3,3)=-\frac{1}{2 \sqrt{2}}&        T_0(2,3,3,5)=-\frac{1}{4}&     T_{-1}(3,3,5,5)=-\frac{1}{2 \sqrt{2}}\\  
T_1(2,3,3,2)=-\frac{1}{2 \sqrt{2}}&        T_0(2,5,3,3)=-\frac{1}{4}&     T_{-1}(3,3,7,7)=\frac{1}{2 \sqrt{2}}\\   
T_1(3,2,2,3)=-\frac{1}{2 \sqrt{2}}&        T_0(3,2,5,3)=-\frac{1}{4}&     T_{-1}(3,5,5,3)=-\frac{1}{2 \sqrt{2}}\\  
T_1(3,3,2,2)=-\frac{1}{2 \sqrt{2}}&        T_0(3,3,2,5)=-\frac{1}{4}&     T_{-1}(3,7,7,3)=\frac{1}{2 \sqrt{2}}\\   
T_1(3,3,6,6)=\frac{1}{2 \sqrt{2}}&         T_0(3,3,5,2)=-\frac{1}{4}&     T_{-1}(5,3,3,5)=-\frac{1}{2 \sqrt{2}}\\  
T_1(3,6,6,3)=\frac{1}{2 \sqrt{2}}&         T_0(3,3,6,7)=\frac{1}{4}&      T_{-1}(5,5,3,3)=-\frac{1}{2 \sqrt{2}}\\  
T_1(6,3,3,6)=\frac{1}{2 \sqrt{2}}&         T_0(3,3,7,6)=\frac{1}{4}&      T_{-1}(7,3,3,7)=\frac{1}{2 \sqrt{2}}\\   
T_1(6,6,3,3)=\frac{1}{2 \sqrt{2}}&         T_0(3,5,2,3)=-\frac{1}{4}&     T_{-1}(7,7,3,3)=\frac{1}{2 \sqrt{2}} \\    
    &                                       T_0(3,6,7,3)=\frac{1}{4}&    \\
    &                                       T_0(3,7,6,3)=\frac{1}{4}&    \\
     &                                      T_0(5,2,3,3)=-\frac{1}{4}&   \\
      &                                     T_0(5,3,3,2)=-\frac{1}{4}&   \\
       &                                    T_0(6,3,3,7)=\frac{1}{4}&  \\  
        &                                   T_0(6,7,3,3)=\frac{1}{4}&  \\  
         &                                  T_0(7,3,3,6)=\frac{1}{4}&   \\ 
          &                                 T_0(7,6,3,3)=\frac{1}{4}&\\    \hline\hline
\caption{$S_6^{(a)}$.}
\label{Table:S6a}
\end{longtable} 

% S6b
\begin{longtable}{LLL}\hline \hline
Sz=+1 & Sz=0  & Sz=-1  \\ \hline\endfirsthead 
\hline Sz=+1 \text{ (cont.)} & Sz=0 \text{ (cont.)}& Sz=-1 \text{ (cont.)} \\ \hline\endhead 
T_1(2,3,2,3)=-\frac{1}{2}&         T_0(2,3,5,3)=-\frac{1}{2 \sqrt{2}}&           T_{-1}(3,5,3,5)=-\frac{1}{2}\\     
T_1(3,2,3,2)=-\frac{1}{2}&         T_0(3,2,3,5)=-\frac{1}{2 \sqrt{2}}&           T_{-1}(3,7,3,7)=\frac{1}{2}\\     
T_1(3,6,3,6)=\frac{1}{2}&          T_0(3,5,3,2)=-\frac{1}{2 \sqrt{2}}&           T_{-1}(5,3,5,3)=-\frac{1}{2}\\    
T_1(6,3,6,3)=\frac{1}{2}&          T_0(3,6,3,7)=\frac{1}{2 \sqrt{2}}&            T_{-1}(7,3,7,3)=\frac{1}{2}    \\   
     &                              T_0(3,7,3,6)=\frac{1}{2 \sqrt{2}}&          \\
      &                             T_0(5,3,2,3)=-\frac{1}{2 \sqrt{2}}&         \\
      &                             T_0(6,3,7,3)=\frac{1}{2 \sqrt{2}}&          \\
      &                             T_0(7,3,6,3)=\frac{1}{2 \sqrt{2}}&\\ \hline \hline
\caption{$S_6^{(b)}$.}
\label{Table:S6b}
\end{longtable} 

% S7a
\begin{longtable}{LLL}\hline \hline
Sz=+1 & Sz=0  & Sz=-1  \\ \hline\endfirsthead 
\hline Sz=+1 \text{ (cont.)} & Sz=0 \text{ (cont.)}& Sz=-1 \text{ (cont.)} \\ \hline\endhead 
T_1(0,2,4,2)=-\frac{1}{2 \sqrt{3}}&        T_0(0,2,4,5)=-\frac{1}{2 \sqrt{6}}&      T_{-1}(0,5,4,5)=-\frac{1}{2 \sqrt{3}}\\      
T_1(0,6,4,6)=\frac{1}{2 \sqrt{3}}&         T_0(0,5,4,2)=-\frac{1}{2 \sqrt{6}}&      T_{-1}(0,7,4,7)=\frac{1}{2 \sqrt{3}}\\    
T_1(1,2,1,2)=\frac{1}{2 \sqrt{3}}&         T_0(0,6,4,7)=\frac{1}{2 \sqrt{6}}&       T_{-1}(1,5,1,5)=\frac{1}{2 \sqrt{3}}\\    
T_1(1,6,1,6)=-\frac{1}{2 \sqrt{3}}&        T_0(0,7,4,6)=\frac{1}{2 \sqrt{6}}&       T_{-1}(1,7,1,7)=-\frac{1}{2 \sqrt{3}}\\   
T_1(2,0,2,4)=-\frac{1}{2 \sqrt{3}}&        T_0(1,2,1,5)=\frac{1}{2 \sqrt{6}}&       T_{-1}(4,5,0,5)=-\frac{1}{2 \sqrt{3}}\\   
T_1(2,1,2,1)=\frac{1}{2 \sqrt{3}}&         T_0(1,5,1,2)=\frac{1}{2 \sqrt{6}}&       T_{-1}(4,7,0,7)=\frac{1}{2 \sqrt{3}}\\    
T_1(2,4,2,0)=-\frac{1}{2 \sqrt{3}}&        T_0(1,6,1,7)=-\frac{1}{2 \sqrt{6}}&      T_{-1}(5,0,5,4)=-\frac{1}{2 \sqrt{3}}\\   
T_1(4,2,0,2)=-\frac{1}{2 \sqrt{3}}&        T_0(1,7,1,6)=-\frac{1}{2 \sqrt{6}}&      T_{-1}(5,1,5,1)=\frac{1}{2 \sqrt{3}}\\
\end{longtable} 

% S7b
\begin{longtable}{LLL}\hline \hline
Sz=+1 & Sz=0  & Sz=-1  \\ \hline\endfirsthead 
\hline Sz=+1 \text{ (cont.)} & Sz=0 \text{ (cont.)}& Sz=-1 \text{ (cont.)} \\ \hline\endhead 
T_1(0,2,1,5)=-\frac{\sqrt{\frac{3}{10}}}{4}&         T_0(0,2,4,5)=-\frac{1}{2 \sqrt{30}}&                  T_{-1}(0,5,4,5)=\frac{1}{4 \sqrt{15}}\\              
T_1(0,2,4,2)=\frac{1}{4 \sqrt{15}}&                  T_0(0,5,1,5)=\frac{\sqrt{\frac{3}{5}}}{4}&            T_{-1}(0,7,4,7)=-\frac{1}{4 \sqrt{15}}\\             
T_1(0,5,0,5)=\frac{\sqrt{\frac{3}{5}}}{2}&           T_0(0,5,4,2)=-\frac{1}{2 \sqrt{30}}&                  T_{-1}(1,2,4,5)=-\frac{\sqrt{\frac{3}{10}}}{4}\\     
T_1(0,5,1,2)=-\frac{\sqrt{\frac{3}{10}}}{4}&         T_0(0,6,4,7)=\frac{1}{2 \sqrt{30}}&                   T_{-1}(1,5,1,5)=\frac{1}{2 \sqrt{15}}\\              
T_1(0,6,1,7)=\frac{\sqrt{\frac{3}{10}}}{4}&          T_0(0,7,1,7)=-\frac{\sqrt{\frac{3}{5}}}{4}&           T_{-1}(1,5,4,2)=-\frac{\sqrt{\frac{3}{10}}}{4}\\     
T_1(0,6,4,6)=-\frac{1}{4 \sqrt{15}}&                 T_0(0,7,4,6)=\frac{1}{2 \sqrt{30}}&                   T_{-1}(1,6,4,7)=\frac{\sqrt{\frac{3}{10}}}{4}\\      
T_1(0,7,0,7)=-\frac{\sqrt{\frac{3}{5}}}{2}&          T_0(1,2,1,5)=-\frac{1}{\sqrt{30}}&                    T_{-1}(1,7,1,7)=-\frac{1}{2 \sqrt{15}}\\             
T_1(0,7,1,6)=\frac{\sqrt{\frac{3}{10}}}{4}&          T_0(1,2,4,2)=\frac{\sqrt{\frac{3}{5}}}{4}&            T_{-1}(1,7,4,6)=\frac{\sqrt{\frac{3}{10}}}{4}\\
\end{longtable} 

% S7c
\begin{longtable}{LLL}\hline \hline
Sz=+1 & Sz=0  & Sz=-1  \\ \hline\endfirsthead 
\hline Sz=+1 \text{ (cont.)} & Sz=0 \text{ (cont.)}& Sz=-1 \text{ (cont.)} \\ \hline\endhead 
T_1(0,0,5,5)=\frac{1}{3 \sqrt{2}}&          T_0(0,1,5,5)=\frac{1}{6 \sqrt{2}}&          T_{-1}(1,1,5,5)=\frac{1}{6 \sqrt{2}}\\    
T_1(0,0,7,7)=-\frac{1}{3 \sqrt{2}}&         T_0(0,1,7,7)=-\frac{1}{6 \sqrt{2}}&         T_{-1}(1,1,7,7)=-\frac{1}{6 \sqrt{2}}\\   
T_1(0,1,2,5)=-\frac{1}{6}&                  T_0(0,4,2,5)=-\frac{1}{6}&                  T_{-1}(1,4,2,5)=-\frac{1}{6}\\            
T_1(0,1,6,7)=\frac{1}{6}&                   T_0(0,4,6,7)=\frac{1}{6}&                   T_{-1}(1,4,6,7)=\frac{1}{6}\\             
T_1(0,5,2,1)=-\frac{1}{6}&                  T_0(0,5,2,4)=-\frac{1}{6}&                  T_{-1}(1,5,2,4)=-\frac{1}{6}\\            
T_1(0,5,5,0)=\frac{1}{3 \sqrt{2}}&          T_0(0,5,5,1)=\frac{1}{6 \sqrt{2}}&          T_{-1}(1,5,5,1)=\frac{1}{6 \sqrt{2}}\\    
T_1(0,7,6,1)=\frac{1}{6}&                   T_0(0,7,6,4)=\frac{1}{6}&                   T_{-1}(1,7,6,4)=\frac{1}{6}\\             
T_1(0,7,7,0)=-\frac{1}{3 \sqrt{2}}&         T_0(0,7,7,1)=-\frac{1}{6 \sqrt{2}}&         T_{-1}(1,7,7,1)=-\frac{1}{6 \sqrt{2}}\\
\end{longtable}

% S7d  
\begin{longtable}{LLL}\hline \hline
Sz=+1 & Sz=0  & Sz=-1  \\ \hline\endfirsthead 
\hline Sz=+1 \text{ (cont.)} & Sz=0 \text{ (cont.)}& Sz=-1 \text{ (cont.)} \\ \hline\endhead 
T_1(0,0,5,5)=\frac{1}{6 \sqrt{13}}&         T_0(0,1,5,5)=\frac{1}{12 \sqrt{13}}&        T_{-1}(0,4,5,5)=\frac{3}{4 \sqrt{13}}\\    
T_1(0,0,7,7)=-\frac{1}{6 \sqrt{13}}&        T_0(0,1,7,7)=-\frac{1}{12 \sqrt{13}}&       T_{-1}(0,4,7,7)=-\frac{3}{4 \sqrt{13}}\\  
T_1(0,1,2,5)=-\frac{1}{6 \sqrt{26}}&        T_0(0,2,5,4)=\frac{3}{4 \sqrt{26}}&         T_{-1}(0,5,5,4)=\frac{3}{4 \sqrt{13}}\\   
T_1(0,1,6,7)=\frac{1}{6 \sqrt{26}}&         T_0(0,4,2,5)=\frac{7}{12 \sqrt{26}}&        T_{-1}(0,7,7,4)=-\frac{3}{4 \sqrt{13}}\\  
T_1(0,2,2,4)=\frac{3}{4 \sqrt{13}}&         T_0(0,4,5,2)=\frac{3}{4 \sqrt{26}}&         T_{-1}(1,1,5,5)=-\frac{2}{3 \sqrt{13}}\\  
T_1(0,4,2,2)=\frac{3}{4 \sqrt{13}}&         T_0(0,4,6,7)=-\frac{7}{12 \sqrt{26}}&       T_{-1}(1,1,7,7)=\frac{2}{3 \sqrt{13}}\\   
T_1(0,4,6,6)=-\frac{3}{4 \sqrt{13}}&        T_0(0,4,7,6)=-\frac{3}{4 \sqrt{26}}&        T_{-1}(1,4,2,5)=-\frac{1}{6 \sqrt{26}}\\  
T_1(0,5,2,1)=-\frac{1}{6 \sqrt{26}}&        T_0(0,5,2,4)=\frac{7}{12 \sqrt{26}}&        T_{-1}(1,4,6,7)=\frac{1}{6 \sqrt{26}}\\
\end{longtable} 

% S7e
\begin{longtable}{LLL}\hline \hline
Sz=+1 & Sz=0  & Sz=-1  \\ \hline\endfirsthead 
\hline Sz=+1 \text{ (cont.)} & Sz=0 \text{ (cont.)}& Sz=-1 \text{ (cont.)} \\ \hline\endhead 
T_1(0,0,5,5)=\frac{3}{2 \sqrt{130}}&                           T_0(0,1,5,5)=\frac{3}{4 \sqrt{130}}&                  T_{-1}(0,4,5,5)=\frac{1}{4 \sqrt{130}}\\               
T_1(0,0,7,7)=-\frac{3}{2 \sqrt{130}}&                          T_0(0,1,7,7)=-\frac{3}{4 \sqrt{130}}&                 T_{-1}(0,4,7,7)=-\frac{1}{4 \sqrt{130}}\\              
T_1(0,1,2,5)=\frac{7}{8 \sqrt{65}}&                            T_0(0,2,5,4)=-\frac{3}{2 \sqrt{65}}&                  T_{-1}(0,5,5,4)=\frac{1}{4 \sqrt{130}}\\               
T_1(0,1,5,2)=-\frac{\sqrt{\frac{13}{5}}}{8}&                   T_0(0,4,2,5)=\frac{1}{\sqrt{65}}&                     T_{-1}(0,7,7,4)=-\frac{1}{4 \sqrt{130}}\\              
T_1(0,1,6,7)=-\frac{7}{8 \sqrt{65}}&                           T_0(0,4,5,2)=-\frac{3}{2 \sqrt{65}}&                  T_{-1}(1,1,5,5)=\frac{1}{2 \sqrt{130}}\\               
T_1(0,1,7,6)=\frac{\sqrt{\frac{13}{5}}}{8}&                    T_0(0,4,6,7)=-\frac{1}{\sqrt{65}}&                    T_{-1}(1,1,7,7)=-\frac{1}{2 \sqrt{130}}\\              
T_1(0,2,2,4)=\frac{1}{4 \sqrt{130}}&                           T_0(0,4,7,6)=\frac{3}{2 \sqrt{65}}&                   T_{-1}(1,2,5,4)=-\frac{\sqrt{\frac{13}{5}}}{8}\\       
T_1(0,2,5,1)=-\frac{\sqrt{\frac{13}{5}}}{8}&                   T_0(0,5,2,4)=\frac{1}{\sqrt{65}}&                     T_{-1}(1,4,2,5)=\frac{7}{8 \sqrt{65}}\\
\end{longtable} 

% G7
\begin{longtable}{LLL}\hline \hline
Sz=+1 & Sz=0  & Sz=-1  \\ \hline\endfirsthead 
\hline Sz=+1 \text{ (cont.)} & Sz=0 \text{ (cont.)}& Sz=-1 \text{ (cont.)} \\ \hline\endhead 
T_1(0,1,2,5)=\frac{1}{4 \sqrt{5}}&            T_0(0,1,5,5)=\frac{1}{2 \sqrt{10}}&           T_{-1}(0,4,5,5)=\frac{1}{2 \sqrt{10}}\\    
T_1(0,1,5,2)=\frac{1}{4 \sqrt{5}}&            T_0(0,1,7,7)=-\frac{1}{2 \sqrt{10}}&          T_{-1}(0,4,7,7)=-\frac{1}{2 \sqrt{10}}\\   
T_1(0,1,6,7)=-\frac{1}{4 \sqrt{5}}&           T_0(0,2,4,5)=-\frac{1}{4 \sqrt{5}}&           T_{-1}(0,5,5,4)=-\frac{1}{2 \sqrt{10}}\\   
T_1(0,1,7,6)=-\frac{1}{4 \sqrt{5}}&           T_0(0,5,4,2)=\frac{1}{4 \sqrt{5}}&            T_{-1}(0,7,7,4)=\frac{1}{2 \sqrt{10}}\\    
T_1(0,2,1,5)=-\frac{1}{4 \sqrt{5}}&           T_0(0,5,5,1)=-\frac{1}{2 \sqrt{10}}&          T_{-1}(1,2,4,5)=-\frac{1}{4 \sqrt{5}}\\    
T_1(0,2,2,4)=\frac{1}{2 \sqrt{10}}&           T_0(0,6,4,7)=\frac{1}{4 \sqrt{5}}&            T_{-1}(1,2,5,4)=\frac{1}{4 \sqrt{5}}\\     
T_1(0,2,5,1)=-\frac{1}{4 \sqrt{5}}&           T_0(0,7,4,6)=-\frac{1}{4 \sqrt{5}}&           T_{-1}(1,4,2,5)=-\frac{1}{4 \sqrt{5}}\\    
T_1(0,4,2,2)=-\frac{1}{2 \sqrt{10}}&          T_0(0,7,7,1)=\frac{1}{2 \sqrt{10}}&           T_{-1}(1,4,5,2)=-\frac{1}{4 \sqrt{5}}\\
\end{longtable}

% S8a
\begin{longtable}{LLL}\hline \hline
Sz=+1 & Sz=0  & Sz=-1  \\ \hline\endfirsthead 
\hline Sz=+1 \text{ (cont.)} & Sz=0 \text{ (cont.)}& Sz=-1 \text{ (cont.)} \\ \hline\endhead 
T_1(0,2,3,7)=-\frac{1}{3 \sqrt{6}}&        T_0(0,3,5,7)=-\frac{1}{4 \sqrt{3}}&      T_{-1}(1,3,5,7)=-\frac{1}{4 \sqrt{3}}\\    
T_1(0,2,7,3)=\frac{1}{6 \sqrt{6}}&         T_0(0,3,7,5)=\frac{1}{4 \sqrt{3}}&       T_{-1}(1,3,7,5)=\frac{1}{4 \sqrt{3}}\\     
T_1(0,3,2,7)=-\frac{1}{3 \sqrt{6}}&        T_0(0,5,7,3)=\frac{1}{4 \sqrt{3}}&       T_{-1}(1,5,7,3)=\frac{1}{4 \sqrt{3}}\\     
T_1(0,3,5,6)=-\frac{1}{6 \sqrt{6}}&        T_0(0,7,5,3)=-\frac{1}{4 \sqrt{3}}&      T_{-1}(1,7,5,3)=-\frac{1}{4 \sqrt{3}}\\    
T_1(0,3,6,5)=\frac{1}{3 \sqrt{6}}&         T_0(1,2,3,7)=-\frac{1}{3 \sqrt{6}}&      T_{-1}(2,3,4,7)=\frac{1}{6 \sqrt{6}}\\     
T_1(0,3,7,2)=\frac{1}{6 \sqrt{6}}&         T_0(1,2,7,3)=-\frac{1}{12 \sqrt{6}}&     T_{-1}(2,3,7,4)=-\frac{1}{3 \sqrt{6}}\\    
T_1(0,5,3,6)=\frac{1}{3 \sqrt{6}}&         T_0(1,3,2,7)=-\frac{1}{12 \sqrt{6}}&     T_{-1}(2,4,3,7)=-\frac{1}{3 \sqrt{6}}\\    
T_1(0,5,6,3)=\frac{1}{3 \sqrt{6}}&         T_0(1,3,5,6)=\frac{1}{12 \sqrt{6}}&      T_{-1}(2,4,7,3)=-\frac{1}{3 \sqrt{6}}\\
\end{longtable}

% S8b
\begin{longtable}{LLL}\hline \hline
Sz=+1 & Sz=0  & Sz=-1  \\ \hline\endfirsthead 
\hline Sz=+1 \text{ (cont.)} & Sz=0 \text{ (cont.)}& Sz=-1 \text{ (cont.)} \\ \hline\endhead 
T_1(0,2,3,7)=-\frac{5}{12 \sqrt{21}}&         T_0(0,3,5,7)=-\frac{1}{4 \sqrt{3}}&      T_{-1}(1,3,5,7)=\frac{1}{2 \sqrt{42}}\\       
T_1(0,2,7,3)=-\frac{11}{12 \sqrt{21}}&        T_0(0,3,7,5)=\frac{1}{4 \sqrt{3}}&       T_{-1}(1,3,7,5)=-\frac{1}{2 \sqrt{42}}\\    
T_1(0,3,2,7)=-\frac{5}{12 \sqrt{21}}&         T_0(0,5,7,3)=\frac{1}{4 \sqrt{3}}&       T_{-1}(1,5,7,3)=-\frac{1}{2 \sqrt{42}}\\    
T_1(0,3,5,6)=\frac{11}{12 \sqrt{21}}&         T_0(0,7,5,3)=-\frac{1}{4 \sqrt{3}}&      T_{-1}(1,7,5,3)=\frac{1}{2 \sqrt{42}}\\     
T_1(0,3,6,5)=\frac{5}{12 \sqrt{21}}&          T_0(1,2,3,7)=-\frac{1}{3 \sqrt{6}}&      T_{-1}(2,3,4,7)=-\frac{11}{12 \sqrt{21}}\\  
T_1(0,3,7,2)=-\frac{11}{12 \sqrt{21}}&        T_0(1,2,7,3)=-\frac{1}{12 \sqrt{6}}&     T_{-1}(2,3,7,4)=-\frac{5}{12 \sqrt{21}}\\   
T_1(0,5,3,6)=\frac{5}{12 \sqrt{21}}&          T_0(1,3,2,7)=-\frac{1}{12 \sqrt{6}}&     T_{-1}(2,4,3,7)=-\frac{5}{12 \sqrt{21}}\\   
T_1(0,5,6,3)=\frac{5}{12 \sqrt{21}}&          T_0(1,3,5,6)=\frac{1}{12 \sqrt{6}}&      T_{-1}(2,4,7,3)=-\frac{5}{12 \sqrt{21}}\\
\end{longtable}

% G8
\begin{longtable}{LLL}\hline \hline
Sz=+1 & Sz=0  & Sz=-1  \\ \hline\endfirsthead 
\hline Sz=+1 \text{ (cont.)} & Sz=0 \text{ (cont.)}& Sz=-1 \text{ (cont.)} \\ \hline\endhead 
T_1(0,2,3,7)=\frac{1}{8}&            T_0(0,5,3,7)=\frac{1}{4 \sqrt{2}}&        T_{-1}(1,5,3,7)=\frac{1}{4 \sqrt{2}}\\                  
T_1(0,2,7,3)=\frac{1}{8}&            T_0(0,7,3,5)=-\frac{1}{4 \sqrt{2}}&       T_{-1}(1,7,3,5)=-\frac{1}{4 \sqrt{2}}\\      
T_1(0,3,2,7)=-\frac{1}{8}&           T_0(1,2,7,3)=\frac{1}{8}&                 T_{-1}(2,3,4,7)=\frac{1}{8}\\                
T_1(0,3,5,6)=\frac{1}{8}&            T_0(1,3,2,7)=-\frac{1}{8}&                T_{-1}(2,3,7,4)=-\frac{1}{8}\\               
T_1(0,3,6,5)=\frac{1}{8}&            T_0(1,3,5,6)=\frac{1}{8}&                 T_{-1}(2,4,3,7)=-\frac{1}{8}\\               
T_1(0,3,7,2)=-\frac{1}{8}&           T_0(1,3,6,5)=\frac{1}{8}&                 T_{-1}(2,4,7,3)=\frac{1}{8}\\                
T_1(0,5,3,6)=\frac{1}{8}&            T_0(1,3,7,2)=-\frac{1}{8}&                T_{-1}(2,7,3,4)=\frac{1}{8}\\                
T_1(0,5,6,3)=-\frac{1}{8}&           T_0(1,5,6,3)=-\frac{1}{8}&                T_{-1}(2,7,4,3)=-\frac{1}{8}\\
\end{longtable}

% S9 
\begin{longtable}{LLL}\hline \hline
Sz=+1 & Sz=0  & Sz=-1  \\ \hline\endfirsthead 
\hline Sz=+1 \text{ (cont.)} & Sz=0 \text{ (cont.)}& Sz=-1 \text{ (cont.)} \\ \hline\endhead 
T_1(2,3,6,8)=-\frac{1}{2 \sqrt{3}}&        T_0(2,3,7,8)=-\frac{1}{2 \sqrt{6}}&     T_{-1}(3,5,7,8)=\frac{1}{4 \sqrt{3}}\\  
T_1(2,3,8,6)=\frac{1}{4 \sqrt{3}}&         T_0(2,3,8,7)=\frac{1}{4 \sqrt{6}}&      T_{-1}(3,5,8,7)=-\frac{1}{2 \sqrt{3}}\\ 
T_1(2,6,3,8)=\frac{1}{4 \sqrt{3}}&         T_0(2,7,3,8)=\frac{1}{4 \sqrt{6}}&      T_{-1}(3,7,5,8)=\frac{1}{4 \sqrt{3}}\\  
T_1(2,6,8,3)=\frac{1}{4 \sqrt{3}}&         T_0(2,7,8,3)=\frac{1}{4 \sqrt{6}}&      T_{-1}(3,7,8,5)=-\frac{1}{2 \sqrt{3}}\\ 
T_1(2,8,3,6)=\frac{1}{4 \sqrt{3}}&         T_0(2,8,3,7)=\frac{1}{4 \sqrt{6}}&      T_{-1}(3,8,5,7)=\frac{1}{4 \sqrt{3}}\\  
T_1(2,8,6,3)=-\frac{1}{2 \sqrt{3}}&        T_0(2,8,7,3)=-\frac{1}{2 \sqrt{6}}&     T_{-1}(3,8,7,5)=\frac{1}{4 \sqrt{3}}\\  
T_1(3,2,6,8)=\frac{1}{4 \sqrt{3}}&         T_0(3,2,7,8)=\frac{1}{4 \sqrt{6}}&      T_{-1}(5,3,7,8)=-\frac{1}{2 \sqrt{3}}\\ 
T_1(3,2,8,6)=-\frac{1}{2 \sqrt{3}}&        T_0(3,2,8,7)=-\frac{1}{2 \sqrt{6}}&     T_{-1}(5,3,8,7)=\frac{1}{4 \sqrt{3}}\\
\end{longtable}

% G9
\begin{longtable}{LLL}\hline \hline
Sz=+1 & Sz=0  & Sz=-1  \\ \hline\endfirsthead 
\hline Sz=+1 \text{ (cont.)} & Sz=0 \text{ (cont.)}& Sz=-1 \text{ (cont.)} \\ \hline\endhead 
T_1(2,3,8,6)=\frac{1}{4}&        T_0(2,3,8,7)=\frac{1}{4 \sqrt{2}}&       T_{-1}(3,5,7,8)=-\frac{1}{4}\\  
T_1(2,6,3,8)=\frac{1}{4}&        T_0(2,7,3,8)=\frac{1}{4 \sqrt{2}}&       T_{-1}(3,7,5,8)=-\frac{1}{4}\\  
T_1(2,6,8,3)=-\frac{1}{4}&       T_0(2,7,8,3)=-\frac{1}{4 \sqrt{2}}&      T_{-1}(3,8,5,7)=\frac{1}{4}\\   
T_1(2,8,3,6)=-\frac{1}{4}&       T_0(2,8,3,7)=-\frac{1}{4 \sqrt{2}}&      T_{-1}(3,8,7,5)=\frac{1}{4}\\   
T_1(3,2,6,8)=-\frac{1}{4}&       T_0(3,2,7,8)=-\frac{1}{4 \sqrt{2}}&      T_{-1}(5,3,8,7)=\frac{1}{4}\\   
T_1(3,6,2,8)=-\frac{1}{4}&       T_0(3,5,6,8)=-\frac{1}{4 \sqrt{2}}&      T_{-1}(5,7,3,8)=\frac{1}{4}\\   
T_1(3,8,2,6)=\frac{1}{4}&        T_0(3,6,5,8)=-\frac{1}{4 \sqrt{2}}&      T_{-1}(5,7,8,3)=-\frac{1}{4}\\  
T_1(3,8,6,2)=\frac{1}{4}&        T_0(3,7,2,8)=-\frac{1}{4 \sqrt{2}}&      T_{-1}(5,8,3,7)=-\frac{1}{4}\\
\end{longtable}

% S10
\begin{longtable}{LLL}\hline \hline
Sz=+1 & Sz=0  & Sz=-1  \\ \hline\endfirsthead 
\hline Sz=+1 \text{ (cont.)} & Sz=0 \text{ (cont.)}& Sz=-1 \text{ (cont.)} \\ \hline\endhead 
T_1(0,2,5,8)=-\frac{1}{2 \sqrt{30}}&        T_0(0,5,5,8)=\frac{1}{4 \sqrt{15}}&                  T_{-1}(1,5,5,8)=\frac{1}{4 \sqrt{15}}\\        
T_1(0,2,8,5)=-\frac{1}{2 \sqrt{30}}&        T_0(0,5,8,5)=-\frac{1}{2 \sqrt{15}}&                 T_{-1}(1,5,8,5)=-\frac{1}{2 \sqrt{15}}\\  
T_1(0,5,2,8)=\frac{1}{\sqrt{30}}&           T_0(0,7,7,8)=\frac{1}{4 \sqrt{15}}&                  T_{-1}(1,7,7,8)=\frac{1}{4 \sqrt{15}}\\   
T_1(0,5,8,2)=-\frac{1}{2 \sqrt{30}}&        T_0(0,7,8,7)=-\frac{1}{2 \sqrt{15}}&                 T_{-1}(1,7,8,7)=-\frac{1}{2 \sqrt{15}}\\  
T_1(0,6,7,8)=-\frac{1}{2 \sqrt{30}}&        T_0(0,8,5,5)=\frac{1}{4 \sqrt{15}}&                  T_{-1}(1,8,5,5)=\frac{1}{4 \sqrt{15}}\\   
T_1(0,6,8,7)=-\frac{1}{2 \sqrt{30}}&        T_0(0,8,7,7)=\frac{1}{4 \sqrt{15}}&                  T_{-1}(1,8,7,7)=\frac{1}{4 \sqrt{15}}\\   
T_1(0,7,6,8)=\frac{1}{\sqrt{30}}&           T_0(1,2,5,8)=-\frac{\sqrt{\frac{3}{10}}}{4}&         T_{-1}(2,4,5,8)=\frac{1}{2 \sqrt{30}}\\   
T_1(0,7,8,6)=-\frac{1}{2 \sqrt{30}}&        T_0(1,5,2,8)=\frac{\sqrt{\frac{3}{10}}}{4}&          T_{-1}(2,4,8,5)=-\frac{1}{\sqrt{30}}\\
\end{longtable}

% G10
\begin{longtable}{LLL}\hline \hline
Sz=+1 & Sz=0  & Sz=-1  \\ \hline\endfirsthead 
\hline Sz=+1 \text{ (cont.)} & Sz=0 \text{ (cont.)}& Sz=-1 \text{ (cont.)} \\ \hline\endhead 
T_1(0,2,5,8)=\frac{1}{2 \sqrt{10}}&        T_0(0,5,5,8)=\frac{1}{4 \sqrt{5}}&         T_{-1}(1,5,5,8)=\frac{1}{4 \sqrt{5}}\\      
T_1(0,2,8,5)=-\frac{1}{2 \sqrt{10}}&       T_0(0,7,7,8)=\frac{1}{4 \sqrt{5}}&         T_{-1}(1,7,7,8)=\frac{1}{4 \sqrt{5}}\\     
T_1(0,5,8,2)=\frac{1}{2 \sqrt{10}}&        T_0(0,8,5,5)=-\frac{1}{4 \sqrt{5}}&        T_{-1}(1,8,5,5)=-\frac{1}{4 \sqrt{5}}\\    
T_1(0,6,7,8)=\frac{1}{2 \sqrt{10}}&        T_0(0,8,7,7)=-\frac{1}{4 \sqrt{5}}&        T_{-1}(1,8,7,7)=-\frac{1}{4 \sqrt{5}}\\    
T_1(0,6,8,7)=-\frac{1}{2 \sqrt{10}}&       T_0(1,2,5,8)=\frac{1}{4 \sqrt{10}}&        T_{-1}(2,4,5,8)=\frac{1}{2 \sqrt{10}}\\    
T_1(0,7,8,6)=\frac{1}{2 \sqrt{10}}&        T_0(1,2,8,5)=-\frac{1}{2 \sqrt{10}}&       T_{-1}(2,5,4,8)=\frac{1}{2 \sqrt{10}}\\    
T_1(0,8,5,2)=-\frac{1}{2 \sqrt{10}}&       T_0(1,5,2,8)=-\frac{1}{4 \sqrt{10}}&       T_{-1}(2,8,4,5)=-\frac{1}{2 \sqrt{10}}\\   
T_1(0,8,7,6)=-\frac{1}{2 \sqrt{10}}&       T_0(1,5,8,2)=\frac{1}{2 \sqrt{10}}&        T_{-1}(2,8,5,4)=-\frac{1}{2 \sqrt{10}}\\
\end{longtable}  
\end{document}